\documentclass[traditabstract]{aa} % for the abstract without structuration 
\usepackage{graphicx}
\usepackage{natbib,twoopt}
\bibpunct{(}{)}{;}{a}{}{,}
\usepackage{txfonts}
\begin{document}
   \title{The NIR structure of the barred galaxy NGC~253 from VISTA\thanks{This work is based on observations taken at the ESO La Silla Paranal Observatory within the VISTA Science Verification Program ID 60.A-9285(A). }}

   \author{E. Iodice
          \inst{1}
          \and
          M. Arnaboldi\inst{2,3}
          \and
          M. Rejkuba\inst{2,4} 
          \and
          M. J. Neeser\inst{2} 
          \and
          L. Greggio\inst{5} 
          \and
          O.A. Gonzalez\inst{6} 
          \and
          M. Irwin\inst{7} 
          \and
          J.P. Emerson\inst{8} 
          }

   \institute{INAF-Astronomical Observatory of Capodimonte, via
     Moiariello 16, I-80131 Naples, Italy
              \email{iodice@na.astro.it}
         \and
             ESO,  Karl-Schwarzschild-Strasse 2, D-85748 Garching, Germany
         \and  
         INAF-Astronomical Observatory of Turin, Strada Osservatorio 20, 
         I-10025 Turin, Italy
         \and  
         Excellence Cluster Universe, Boltzmannstr. 2, D-85748, Garching, Germany
         \and
         INAF Astronomical Observatory of Padua, Vicolo dell'Osservatorio 5, 
         I-35122 Padova, Italy
         \and
             ESO,  Ave. Alonso de Cordova 3107, Casilla 19, 
             Santiago 19001, Chile
         \and
             Institute of Astronomy, Madingley Road, Cambridge CB03 0HA, UK
             \and
             Astronomy Unit, School of Physics and Astronomy, Queen Mary 
             University of London, Mile End Road, London, E1 4NS, UK
             }

   \date{Received 2014 January 21; accepted 2014 May 26}

\abstract{ 
% context {} 
%{The presence of a bar affects the distribution and dynamics of a stellar disk at all scales, from a fraction of a kpc in the inner central region to tens of kpc at the disk's edge. The quantitative  study of the disk response to a bar can be hampered by the  presence of dust, which is common in late type spirals.}
 % aims {} 
We want to quantify the structures in the stellar disk of the barred Sc galaxy NGC 253 located in the Sculptor group, at 3.47~Mpc
distance.
 % methods {} 
We used J and Ks band images acquired with the VISTA telescope as part of the science verification. The wide field of view and the high angular resolution of this survey facility allow  mapping  the large and small scale structure of the stellar disk in NGC~253. 
We used unsharp masking and two-dimensional modeling of the smooth light distribution in the disk to identify and
measure the substructures induced by the bar in the stellar disk of NGC~253. We built azimuthally-averaged profiles in the J and Ks bands to 
measure the radial surface brightness profile of the central bulge, bar and disk.
% results {}
Moving outward from the galaxy center, we find a nuclear ring within the bright 1~kpc diameter nucleus, then a bar, a ring with 2.9 kpc radius, and spiral arms in the outer disk. From the Ks image we obtain a new measure of the   deprojected  length of the bar of 2.5~kpc. The bar's strength, as derived from the curvature of the dust lanes in the J-Ks image,  is typical of weak bars with  $\Delta \alpha=25$ degree/kpc. From the deprojected  length of the bar, we establish the corotation radius ($R_{CR}=3$~kpc) and bar pattern speed ($\Omega_b = 61.3$ kms$^{-1}$~kpc$^{-1}$), which provides the connection between the  high-frequency structures in the disk and the orbital resonances induced by the bar. The nuclear ring is located at the inner Lindblad resonance. The second ring (at 2.9~kpc) does not have a resonant origin, but it  could be a merger remnant or a transient structure formed during an intermediate stage of the bar formation. The inferred bar pattern speed places the outer Lindblad resonance  within the optical disk at 4.9~kpc, in the same radial  range as the peak in the HI surface density. The disk of NGC~253 has a down-bending profile with a break  at $R \sim$9.3~kpc, which corresponds to about 3 times the scale length of the inner disk. We discuss the evidence for a threshold in  star formation efficiency as a possible explanation of the steep  gradient in the surface brightness profile at large radii.
% conclusions {}
The NIR photometry unveils the dynamical response of the NGC~253  stellar disk to its central bar. The formation of the bar may be
  related to the merger event that determined the truncation of stars  and gas at large radii and the perturbation of the disk's outer edge.}
% 5 {} token are mandatory


\authorrunning{Iodice et al.}
\titlerunning{The NIR structure of NGC~253}
   \maketitle
%
%________________________________________________________________

\section{Introduction}\label{intro}

 NGC~253 is a southern\footnote{RA(J2000)=00h 47m 33s;
   DEC(J2000)=-25d 17m 18s}, barred, edge-on   ($i \simeq$ 74~degrees), 
 spiral galaxy in the Sculptor group at a distance of 3.47 Mpc \citep{RS11}, 
   which yields an image scale of 16.8 parsecs
 per arcsecond ($\sim$1~kpc/arcminute). It is one of
 the best nearby examples of a nuclear starburst galaxy. Even if its
 overall gas and stellar morphology is typical of a spiral galaxy,
 several photometric and kinematical studies on this object have
 revealed that NGC~253 has a rather complicated structure. The deep
 image of \citet{MH97}, reaching 28 mag/arcsec$^2$, shows the presence
 of an extended, asymmetrical stellar halo   with a semi-major axis 
 radius of about 34~kpc, plus a southern spur. The
 stellar disk is much more extended than the HI disk \citep{Boo05},
 contrary to what is normally observed in spiral  galaxies. Furthermore, the 
 HI distribution in NGC~253 presents two
 other features. The HI disk is  less extended on the NE side with respect to
 the SW, and on the same side, a plume is observed which is elongated
 perpendicular to the disk major axis and extends for about 12 kpc.
 This HI plume borders the X-ray halo emission \citep{Pie00} and the
 $H\alpha$ emission \citep{Hoo96} on their northern side. Given the
 spatial connection, such a feature has been related to the central
 starburst or, alternatively, to a minor merger and a gas accretion
 event \citep{Boo05}.  
 
Previous photometric studies in the near infrared ($1-2$~$\mu m$)   revealed the presence of a bar extending 150~arcsec from the   nucleus \citep{Sco85, Forbes92}, in addition to strong nuclear emission. These studies are confined within 3~arcmin radius   ($\sim 3$~kpc) from the center and do not cover the whole disk that extends out to 30~arcmin ($\sim 30$~kpc).

 Toward the nuclear regions, an intense starburst is powering the
 observed outflow of expanding gas shells along the minor axis
 \citep{SW92}. Recently, extraplanar molecular gas was detected by
 ALMA \citep{ALMA} which closely tracks the $H\alpha$ emission. The
 nuclear outflow is also responsible for the extended X-ray plume \citep{FB84}.

 The $H\alpha$ rotation curve along the disk major axis \citep{AM95}
 is asymmetric inside 100 arcsec   ($\sim$1.7~kpc) from the center
 and the steep velocity gradient for $R\le10$ arcsec on the NE side
 suggests the presence of a nuclear ring, which may be responsible for
 the gas supply to the nuclear starburst. From this analysis of the
 bar dynamics one expects an Inner Lindblad Resonance (ILR) at the
 scale of the observed nuclear ring \citep{AM95}.  A nuclear ring of a
 comparable size has been detected by \citet{MS10} based on SINFONI
 photometry and 2D kinematics in the Ks band.  Both kinematical
 studies cited above have shown that there is an offset between the
 kinematic center and the brightest location in the nuclear region of
 NGC~253 \citep[see Fig.1b and Fig.5 in][respectively]{AM95, MS10}.  The puzzle
 of the nucleus in NGC~253 was discussed in detail by \citet{MS10}: the
 SINFONI data have revealed that the IR peak is at about 2.6 arcsec
 away from the center of the 2D velocity map, while it seems to be
 consistent with the location of the strongest compact radio source
 TH2.  This radio source has no optical, IR or X-ray counterpart that
 led to it being excluded as an AGN. Alternatively, since the
 kinematic center is very close to TH2, they suggested the presence of
 a dormant black hole in the center of NGC~253, like SgrA* in the MW.

Taking into account that NGC~253 is a nearby extended object, one
limitation of all previous imaging data is the absence of high angular
resolution covering the entire extent of the galaxy in a single
image. This fact has hampered the study of the fine sub-substructures
and the ability to correlate them with the outer disk and halo. As we
shall discuss in detail in the next sections, this issue is overcome
thanks to the advent of the new generation Wide-Field Imaging (WFI)
cameras. In fact, NGC~253 has been the target of the Science
Verification (SV) for the new ESO survey telescopes VST and VISTA.
The primary goal of SV is to test the expected performance of the
telescope, camera, and of the data reduction pipeline.  NGC~253 was
chosen as a SV target for several reasons. First, its extent fills
most of the VISTA and VST field such that one can check for possible
reflections within the camera optics, and establish most suitable 
techniques for background subtraction. 
Second, as NGC~253 is very dusty, NIR imaging is
a requisite for studying the underlying structure of the disk. And,
finally, a wealth of data is available in the ESO archive (narrow band
$H\alpha$, broad bands from ESO/MPI-2.2WFI, imaging and spectra of the
nucleus from SINFONI at ESO/VLT).  The main scientific goals of the SV
extragalactic mini-survey\footnote{The full SV proposal is available
  at http://www.eso.org/sci/activities/vltsv/vista/index.html} are: 1)
detecting the Red Giant Branch stars in the faint outer halo, by using
the deep exposures, and 2) study of the disk and bulge structure with
shallow exposures. The former science case is presented in \citet{Greggio}. 
In this paper, we focus on the latter science
case and we show the major results on the structure of NGC~253 derived
by the VISTA data in the NIR J and Ks bands. In particular, we 
derived new and more accurate estimates for the bar length and
strength, and discuss the connection between the observed
features in the disk and the Lindblad resonances predicted by the
bar/disk kinematics.

This paper is structured as follows: in Sec.~\ref{data} we present the
observations and data reduction; in Sec.~\ref{morph} we describe the
morphology of NGC~253 in the J and Ks bands; in Sec.~\ref{phot} and
Sec.~\ref{galfit} we carry out the surface photometry
and the two-dimensional model of the light distribution for the whole
system, respectively. Results are discussed in Sec.~\ref{result} and concluding
remarks are drawn in Sec.~\ref{concl}. 

%Throughout this paper we assume a distance of 3.47 Mpc, which yields an image scale of 16.8 parsecs per arcsecond ($\sim$1~kpc/arcminute).

%__________________________________________________________________

\section{Observations \& data reduction }\label{data}
The {\it Visible and Infrared Survey Telescope for Astronomy (VISTA)}
\citep{Em04, Em10}, located at the Paranal Observatory, in Chile, is a
4 meter telescope equipped with the wide-field, near-infrared camera
VIRCAM \citep{da04}. This instrument consists of 16 $2048 \times 2048
$ Raytheon VIRGO HgCdTe detectors non-contiguously covering a $1.29
\times 1.02$ deg$^2$ FoV, in a wavelength range from 0.85 to 2.4
micron.   Hence a single VIRCAM exposure, the so-called {\it
    pawprint}, only covers a FOV of 0.6 deg$^2$.  This is due to the
large gaps between the VIRCAM detectors \citep[$90\%$ and $42.5\%$ of
the x and y axes, respectively, see][]{Em04}.
  The contiguous area of 1.65 deg$^2$, the {\it tile}, is obtained
  by combining a minimum of 6 offset pawprints that, in unit of
  detector size, are 0.475 twice, along the y axis, and 0.95
  once, along the x axis  \citep[see Fig.~5 in][]{Em04}. The mean pixel scale is 0.34~arcsec/pixel.
In order to account for the variable sky fluctuations and bad pixels,
several images are taken by offsetting the telescope in right
ascension and declination ({\it jitter}) and the series of $DIT \times
NDIT$ is repeated at each jitter position, where the Detector
Integration Time (DIT) is a short exposure on the target. 
  The jitter offsets are $\sim20$~arcsec in size (always $< 30$~arcsec).

For NGC~253, the observations were collected in October 2009 and they
consist of two datasets: the {\it deep data}, taken with J, Z and  
  NB118\footnote{NB118 is a narrow band filter centered close to
    1.18micron. See VIRCAM user manual on
    http://www.eso.org/sci/facilities/paranal/instruments/vircam/doc/
    and \citet{MilJen2013}.}  filters and the {\it shallow data}
taken with all broad-band filters   i.e., Z, Y, J, H, Ks.  Deep
data in the J and Z bands are presented in \citet{Greggio}, where the resolved stellar population is discussed. Here we present and analyze the surface photometry in the J and Ks bands.

  The observing strategy adopted for shallow data consisted of
  positioning the galaxy at the center of the pawprint field, 
  almost filling the central detectors, for three exposures in a six pointings tile sequence,
 and positioning the galaxy in the  gaps between VIRCAM detectors in the other three pointings. In all
  pointings, the major axis of the galaxy is aligned parallel to the  short side of a tile.
  Taking into account that gaps between detectors are $\sim4.5$~arcmin in one direction and $\sim10$~arcmin in the other  \citep[see Fig.5 in][]{Em04}, by positioning
  the galaxy in the gap, we obtain an offset-sky exposure. The jitter
  sequences of these frames were used to create the sky frame from
  their median combination.  Before combining them, the offset-sky
  exposures are scaled to a reference frame. The final sky image is
  used for background subtraction of each pointing.
The observing log for J and Ks shallow data of NGC~253 is listed in Tab.~\ref{obslog}.

  As described also in \citet{Greggio}, the data reduction of 
  shallow and deep datasets is carried out using the dedicated CASU
  pipeline, developed specifically for the reduction of the VISTA  data \citep{Irw04}.

\begin{table}
\caption{The observing log for the J and Ks data of NGC~253.}             % title of Table
\label{obslog}      % is used to refer this table in the text
\centering                          % used for centering table
\begin{tabular}{c c c c c}        % centered columns (4 columns)
\hline\hline                 % inserts double horizontal lines
Filter & $DIT \times NDIT$ & Jitter & Tot. Exp. time  & seeing \\  
 & (seconds) &  & (hours) & (arcsec) \\  
\hline                        % inserts single horizontal line
 J & $10 \times 6$ & 24 & 0.6 & 1.1 \\   
 Ks & $12 \times 6$ & 24 & 0.72 & 0.9\\      
\hline                                   %inserts single line
\end{tabular}
\end{table}

\subsection{Integrated magnitudes and limits of the VISTA data}\label{2mass}

\begin{figure}[!ht] 
%\centering
\includegraphics[width=9cm]{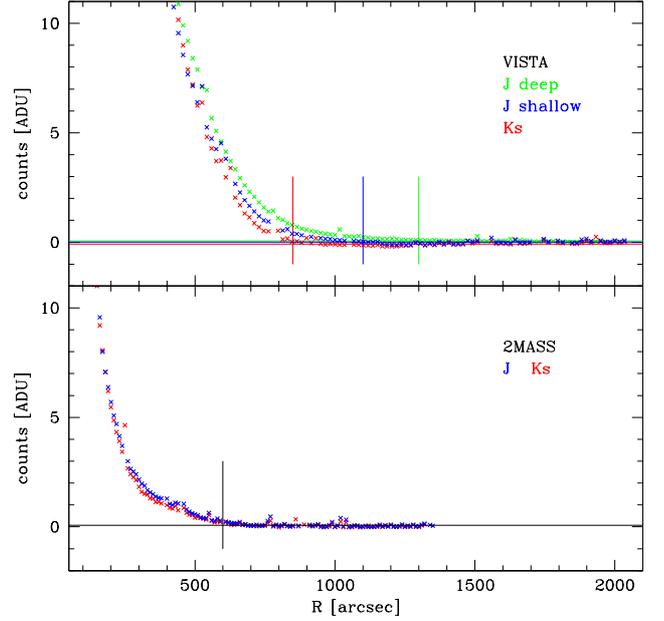}
\caption{ Azimuthally-averaged intensity profiles (counts) as a function of the
  semi-major axis for the VISTA (top panel) and 2MASS data of NGC~253,
  in both J (blue points) and Ks (red points) bands. For the VISTA
  data, the intensity profile is also derived for the deep J band image
  (green points). The vertical lines indicate the outer radii
  corresponding to the ellipse of the limiting surface brightnesses
  for the imaging data in each band. They are $R=600$~arcsec for the
  2MASS images in both J and Ks bands (black vertical line in the
  bottom panel), and $R=1034$~arcsec and $R=830$~arcsec, for the
  shallow VISTA data (top panel) in the J (blue vertical line) and Ks
  (red vertical line) bands respectively, while for the deep J band
  image it is at $R=1305$~arcsec (green vertical line). The horizontal
  lines indicate the residual counts in the background, which are
  $\sim 0.08$ in the 2MASS images (black horizontal line), and $\sim
  -0.08$ in the Ks image (red horizontal line), $\sim 0.002$ and $\sim
  0.06$ in the J shallow and deep images (blue and green horizontal
  lines) for the VISTA data.}
              \label{sky}
   \end{figure}

In order to quantify the limiting surface brightness of the new VISTA
data for NGC~253 we adopted the method described by \citet{PhT06}. On
the sky-subtracted tiles for the shallow J, Ks and deep J imaging data, we extracted the azimuthally-averaged intensity profile  with the IRAF task ELLIPSE. The major axis of the ellipses increases linearly with a step of 50 pixels out to
the edges of the frame. The Position Angle (P.A.) and ellipticity
($\epsilon$) of the ellipses are fixed at a $P.A.= 52\deg$ and $\epsilon
= 0.8$, which are the disk's average  values for $R \ge 200$~arcsec in
NGC~253. In Fig.~\ref{sky}, we show the intensity profile as a function of the semi-major axis
for the VISTA data (top panel) and for the 2MASS data (bottom panel)
for NGC~253. From these intensity profiles, we estimated the distance
from the center where the galaxy's light blends into the background 
 at zero counts per pixel on average. This radius sets the surface
brightness limit of the VISTA and 2MASS photometry. In the J and Ks
2MASS images, this limit is at $R=600$~arcsec, corresponding to a  
limiting surface brightness of $\mu_J = 21.50$~mag arcsec$^{-2}$ and
$\mu_{Ks} = 19.05$~mag arcsec$^{-2}$, respectively. 
The outer limits for the VISTA shallow J and Ks
images are at $R=1034$~arcsec and $R=830$~arcsec, respectively, while
for the deep J band image it is at $R=1305$~arcsec. 
The limiting magnitudes corresponding to these radii are $\mu_J = 23.0 \pm 0.4$~mag~arcsec$^{-2}$ and $\mu_{Ks} = 22.6 \pm 0.6$~mag arcsec$^{-2}$ for the shallow data, and $\mu_J = 25 \pm 1$~mag arcsec$^{-2}$ for the deep J band image. The error estimates on the above quantities take the uncertainties on the photometric calibration ($\sim
0.01$~mag) and sky subtraction ($\sim 0.1$~ADU) into account.

We also measured the integrated magnitudes in two circular apertures
centered on NGC~253. The first aperture is within 300~arcsec, for both
VISTA and 2MASS J and Ks images; the second aperture corresponds to
the outer limit of the J deep and Ks VISTA data derived above. Values 
are listed in Table~\ref{mag}. When the 2MASS magnitudes are
transformed into the VISTA system\footnote{To compare 2MASS magnitudes
  with VISTA ones we have applied the following transformation between
  2MASS and VISTA systems:   $m_{J}^{VISTA} = m_{J}^{2MASS} -0.065
    \left[m_{J}^{2MASS} - m_{Ks}^{2MASS}\right]$, $m_{Ks}^{VISTA} =
    m_{Ks}^{2MASS} +0.01 \left[m_{J}^{2MASS} - m_{Ks}^{2MASS}\right]$
  See the following link
  http://casu.ast.cam.ac.uk/surveys-projects/vista/technical/photometric-properties.}
the magnitudes inside 300~arcsec are consistent within the photometric
errors, in both J and Ks bands.

\begin{table*}
\caption{Integrated magnitudes in circular apertures in the J and
  Ks-bands for 2MASS and VISTA data of NGC~253.}  
  \centering
\begin{tabular}{c c c c c c c}     % 7 columns 
\hline\hline       
Radius & $m_J$ (2MASS) & $m^c_J$ (2MASS) & $m_J$ (VISTA) & $m_{Ks}$ (2MASS) & $m^c_{Ks}$ (2MASS) & $m_{Ks} $(VISTA) \\
 & $\pm 0.01$ & & $\pm 0.011$ & $\pm 0.01$ &  & $\pm 0.011$\\  
(1) & (2) & (3) & (4) & (5) & (6) & (7)\\
\hline                    
%15 & 8.313 & 8.209 & 8.210 & 6.713 & 6.729 & 6.849\\  
300 & 5.08 & 5.01 & 4.986 & 4.00 & 4.01 & 4.001\\ 
830 & & & & & & 3.93\\ 
%1034 & & & & & & \\ 
1305 & & & 4.69 & & & \\ 
\hline                  
\end{tabular}
\tablefoot{Values are not corrected for galactic extinction, since it
  is very low in the NIR J and Ks bands ($A_{\lambda} [J] = 0.013$ mag
  and $A_{\lambda} [Ks] = 0.006$ mag, \citep{SF2011}. Col.~1: Radius
  of the circular aperture in arcsec. Col.~2: Integrated magnitudes in
  the 2MASS J-band. Col.~3: Integrated magnitudes in the J band
  converted from the 2MASS to the VISTA system. Col.~4: Integrated
  magnitudes in the VISTA J-band. Col.~5: Integrated magnitudes in the
  2MASS Ks-band. Col.~6: Integrated magnitudes in the Ks-band
  converted from the 2MASS to the VISTA system. Col.~7: Integrated
  magnitudes in the VISTA Ks-band.}
\label{mag}
\end{table*}

   \section{Morphology of NGC~253 in the NIR J and Ks bands
   }\label{morph}

\begin{figure*}[!ht] 
\centering
\includegraphics[width=14cm]{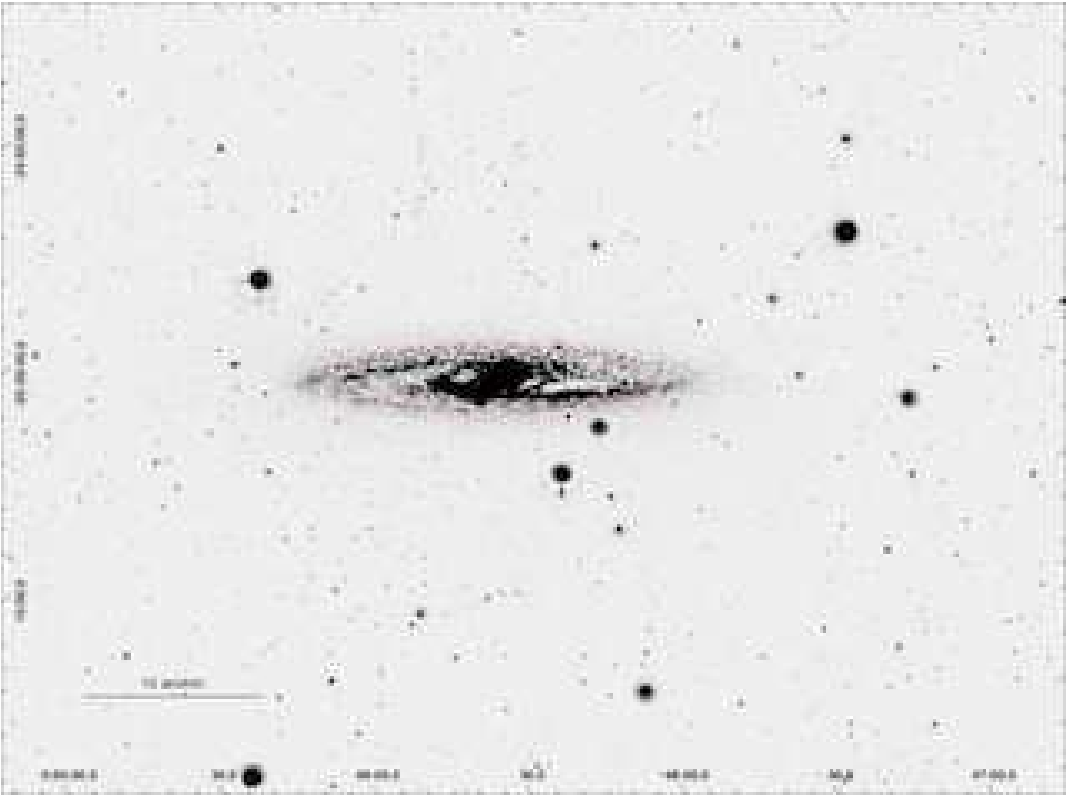}
\caption{ J-band VISTA image of NGC~253 from deep exposures. The
  red dashed line correspond to the ellipses at the break radius of
  the disk (see Sec.~\ref{phot} for details). The image size is $59
  \times 44.2$~arcmin, which corresponds to about 60\% of the whole
  VISTA contiguous area of 1.65 deg$^2$. The full VISTA field of view is shown in Fig.~1 of \citet{Greggio}.}
             \label{tileJ}
   \end{figure*}

\begin{figure*} [!ht] 
\centering
\includegraphics[width=14cm]{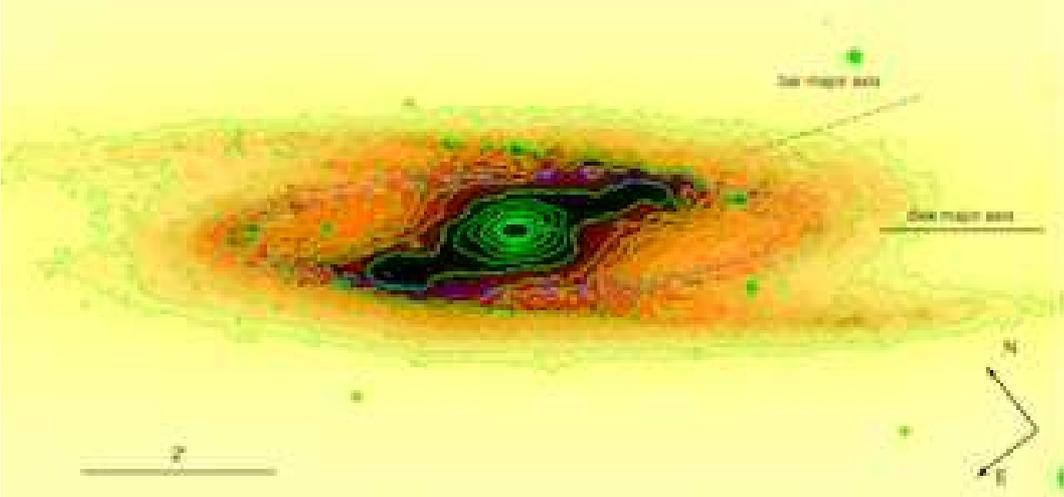}
\caption{The central $11 \times 5$ arcmin region of the Ks band VISTA
  image of NGC~253. The green lines are the isophote contours (the
  outer and inner contour levels correspond to a surface brightness of
  $\mu_{Ks} = 17.6$ mag arcsec$^{-2}$ and $\mu_{Ks} = 13.5$
  mag arcsec$^{-2}$, respectively). On the right side of the image, the
  solid and dashed black lines indicate the directions of the disk and
  bar major axis respectively. The dashed blue annulus encircles the
  ring-like structure   enclosing the bar.}
              \label{tileKs}
   \end{figure*}

   {\it Disk structure: bar, ring and spiral arms -} The morphology of
   the Sculptor Galaxy NGC~253 changes dramatically from optical to
   near-infrared wavelengths\footnote{For visualisation purposes, look
     at the video that cross fades between VISTA
     and optical images of NGC253, available at the following link:
     http://www.eso.org/public/videos/eso1025b/ }.  In the optical,
   the galaxy structure resembles that of an Sc spiral: the disk is
   very dusty and star formation regions dominate the spiral arms
   \citep[see][]{Iod12}. In the J and Ks images of NGC~253, taken at
   VISTA, the most prominent features of the galaxy are
   {\it i)} the bright and almost round nucleus with a diameter of
   about   1~arcmin ($\sim$1~kpc); {\it ii)} the bar, with a
   typical peanut shape ending with very bright edges; {\it iii)}  
     a ring-like structure, located in the main disk, enclosing the
     bar, and {\it iv)} the spiral arms, which start at the end of the
     bar, and dominate the disk. Such structures are already evident in the J-band image
   (Fig.~\ref{tileJ}) but become clearer in the Ks image (see
   Fig.~\ref{tileKs} and Fig.~\ref{imaK_dp}), because of weaker
   dust absorption.

\begin{figure*}[!ht] 
\centering
\includegraphics[width=14cm]{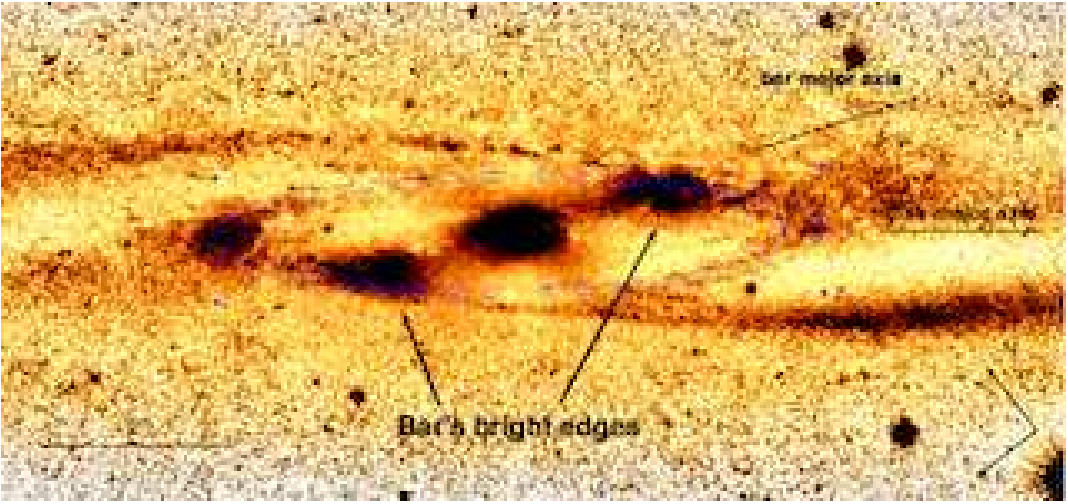}
\caption{The central $11 \times 5$ arcmin region of the unsharped
  masked image of NGC~253 (ratio of the Ks band image and its FMEDIAN
  smoothed version). On the right side of the image, the solid and
  dashed black lines indicate the directions of the disk and bar major
  axis respectively. The dashed blue annulus encircles the ring-like
  structure   enclosing the bar. The two arrows point to the
  bright edges of the bar.}
              \label{FM}
   \end{figure*}

   The multi-component structure of the NGC~253 disk is emphasized by
   the unsharp masked Ks-band image, shown in Fig.~\ref{FM}. It is
   obtained by using the {\it FMEDIAN} task in IRAF, with a smoothing
   box of $150 \times 150$ pixels, and taking the ratio of the Ks-band
   image to its {\it FMEDIAN} smoothed version. In particular, the
   ring-like structure at the end of the bar can be seen. It has a
   radius of about 180~arcsec ($\sim$3~kpc) and appears very bright
   in the SE and NW sides, close to the edges of the bar.

\begin{figure*}[!ht] 
\centering
\includegraphics[width=13cm]{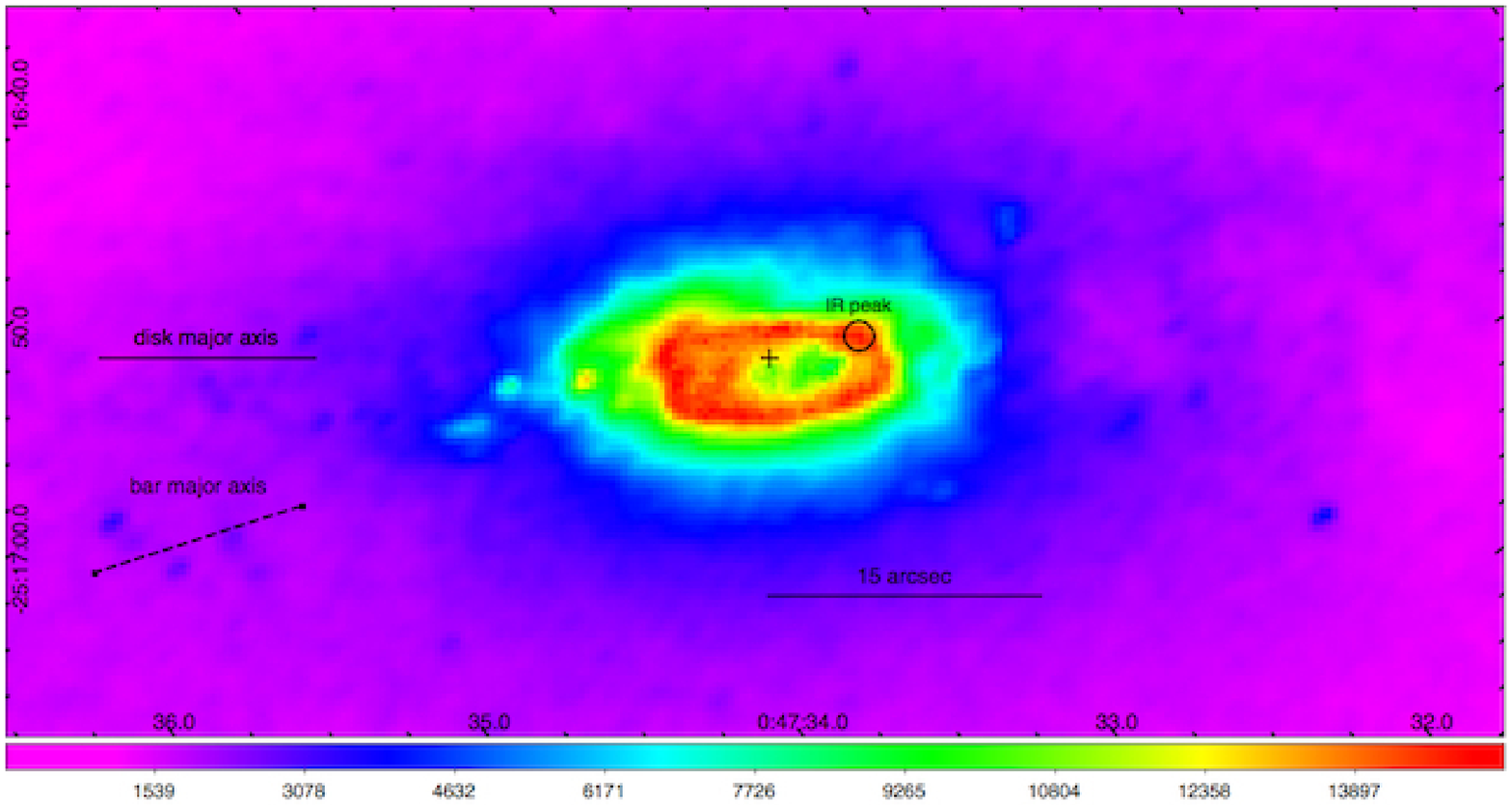}
\caption{Enlargement $80 \times 40$ arcsec of the nuclear region of
  NGC~253 in the Ks image. Orientation is the same as in
  Fig.~\ref{tileKs}, and on the X and Y axis are also indicated the RA
  and DEC coordinates. The black cross is the position of the
  kinematic center derived by \citet{MS10}, at $\alpha = 00^h 47^m
  33.17^s$ and $\delta = -25^{\circ} 17' 17.1''$ (see Sec.\ref{morph}
  for details. The location of the brightest peak of light (IR peak)
  is indicated by the black circle, which is on the SW side at
  $\sim$5.5~arcsec far from the kinematic center.  On the left side of
  the image, the straight and dashed black lines correspond to the
  directions of the disk and bar major axis respectively. }
              \label{zoom}
   \end{figure*}

   {\it The nuclear region - } The zoomed view of  the nuclear region of
   NGC~253 given in Fig.~\ref{zoom} reveals the presence of a   nuclear
     ring of about 30 arcsec diameter ($\sim$0.5~kpc). This feature
   was already detected by \citet{MS10} in the Ks data obtained with
   SINFONI at the VLT. Its morphology and extension are consistent
   with those derived by the Ks VISTA image, with the latter providing
   additional evidence for several luminous peaks distributed along
   this structure, which generate a clumpy azimuthally light
   distribution. Furthermore, as already found by \citet{MS10}, the Ks
   VISTA image confirms that the brightest peak is not coincident with
   the kinematic center, and that it is located on the SW, at 5.5
   arcsec far from the kinematic center (both are marked in
   Fig.~\ref{zoom}). This value has been derived by computing a
   statistic on the Ks image (with the {\it IMEXAM} task in IRAF)
   inside the central 30 arcsec area.

%------------------------------------------------------------------

\section{Surface Photometry }\label{phot}

In this section we describe the surface photometry and the
two-dimensional model of the light distribution for the whole
system. As already mentioned in Sec.~\ref{intro} and Sec.~\ref{morph},
previous kinematical studies showed that the brightest location in
the nuclear region does not coincide with the kinematic center of the
galaxy.  For the following analysis, we adopted as
center of the galaxy the kinematic center found by \citet{MS10} at
$\alpha = 00^h 47^m 33.17^s$ and $\delta = -25^{\circ} 17' 17.1''$,
shown in Fig.~\ref{zoom},   which is consistent with the center of
  the nuclear ring.

\subsection{Fit of the isophotes }\label{ellfit}

\begin{figure*}[!ht] 
%\centering
\includegraphics[width=9cm]{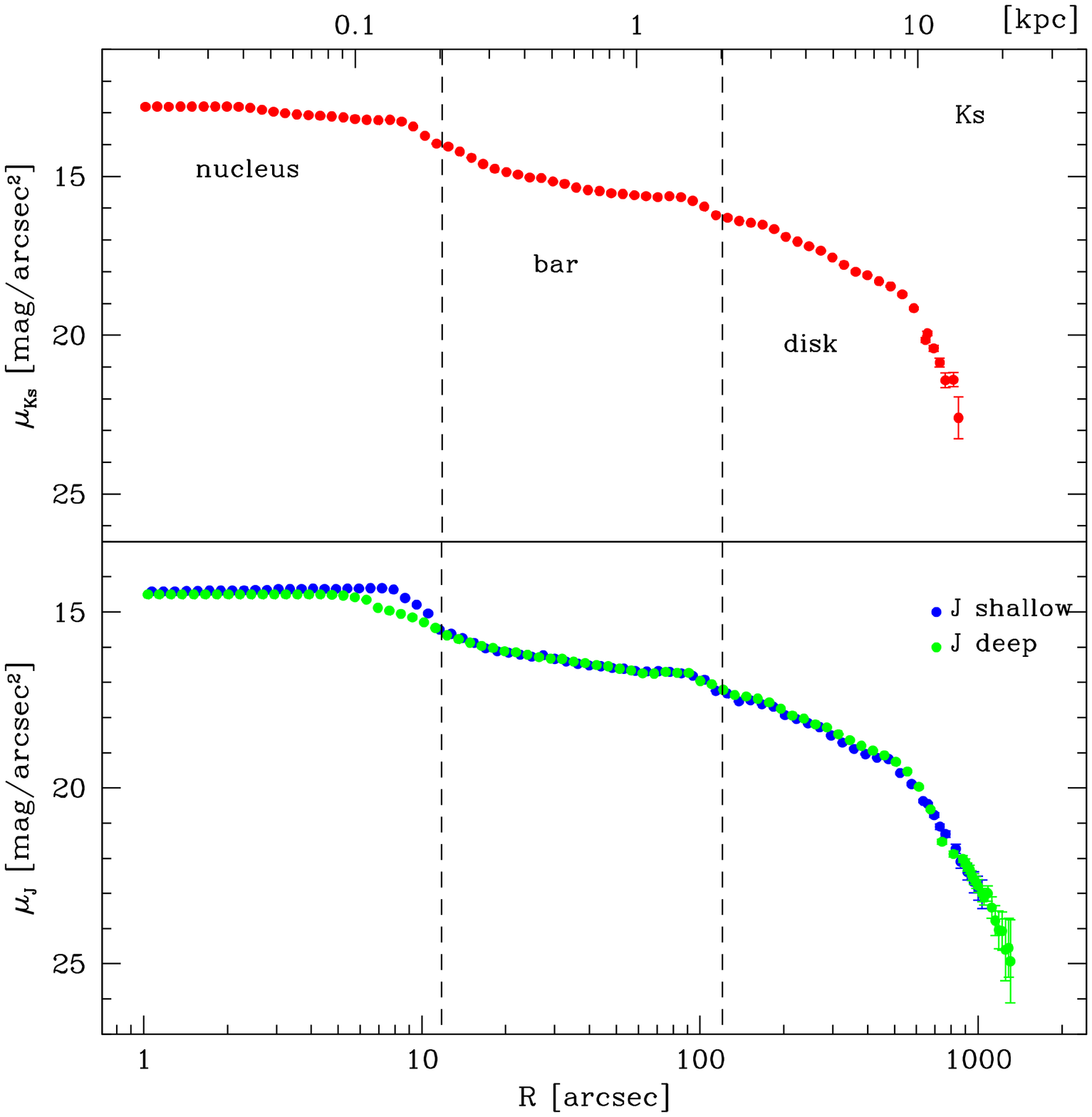}
\includegraphics[width=9cm]{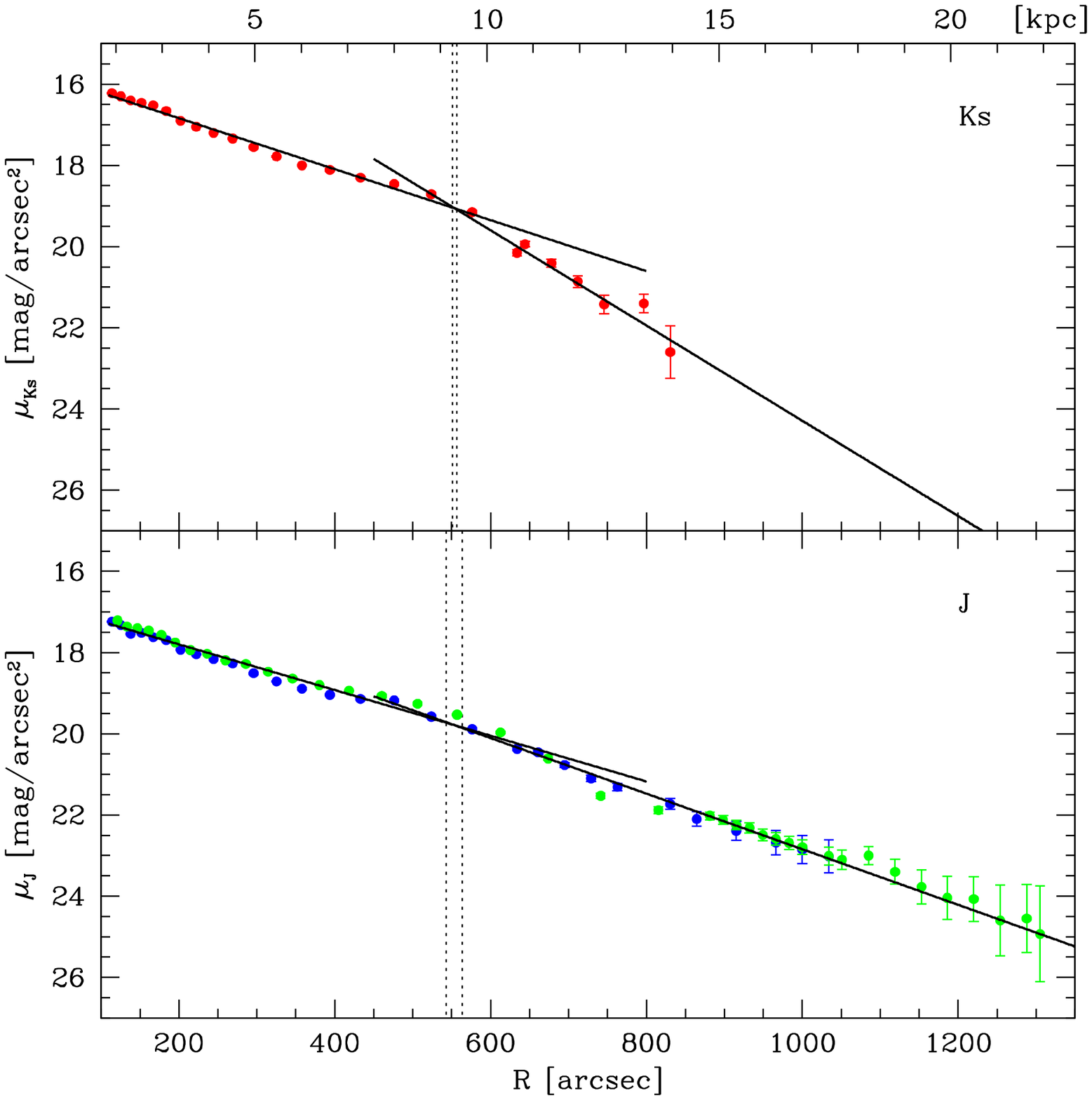}
\caption{Left panel: azimuthally averaged surface brightness
    profiles as function of log(R), with $R$ being the isophote major
    axis from ELLIPSE. Data are for the shallow images in the J (blue
    dots) and Ks-band (red dots) as well as the deep data in J-band
    (green dots). The dashed lines delimit the regions where
    the main components of the galaxy structure are located. Right
    panel: enlargement of the azimuthally averaged surface brightness
    profiles on the disk region in J (bottom) and Ks (top) bands, as a 
    function of the semi-major axis R. The vertical dotted lines indicate
    the range of radii of the break; the straight lines are the best
    fit using an exponential law (see text for details).}
              \label{fit_log}
   \end{figure*}

We used the {\it ELLIPSE} task in IRAF to perform the isophotal
analysis of NGC~253.   The average surface brightness profiles in J
  and Ks shallow data, as well as that derived from the deep J band
  image, are shown in Fig.~\ref{fit_log}.

  {\it Disk structure: bar, ring and spiral arms -} According to the
  outer limits of the surface photometry in the VISTA images, derived
  in Sec.~\ref{2mass}, the average surface brightness profiles extend
  out to 830~arcsec ($\sim$14~kpc) in the Ks band (see
  Fig.~\ref{fit_log} top panel), and out to 1034~arcsec ($\sim$17~kpc)
  and 1305~arcsec ($\sim$22~kpc) in the J band from shallow and deep
  images, respectively (see Fig.~\ref{fit_log} bottom panel).

  The main features are: {\it i)} inner flat profiles for $R \le 12$~arcsec ($\sim$0.2~kpc); {\it ii)} the
  bulge plus bar plateau, $12 \le R \le 120$ arcsec ($0.2 \le R \le  2$~kpc); 
  {\it iii)} the very extended exponential disk for $R\ge 120$~arcsec ($R \ge 2$~kpc) (see Fig.~\ref{fit_log}).

\begin{figure*}[!ht] 
%\centring
\includegraphics[width=9cm]{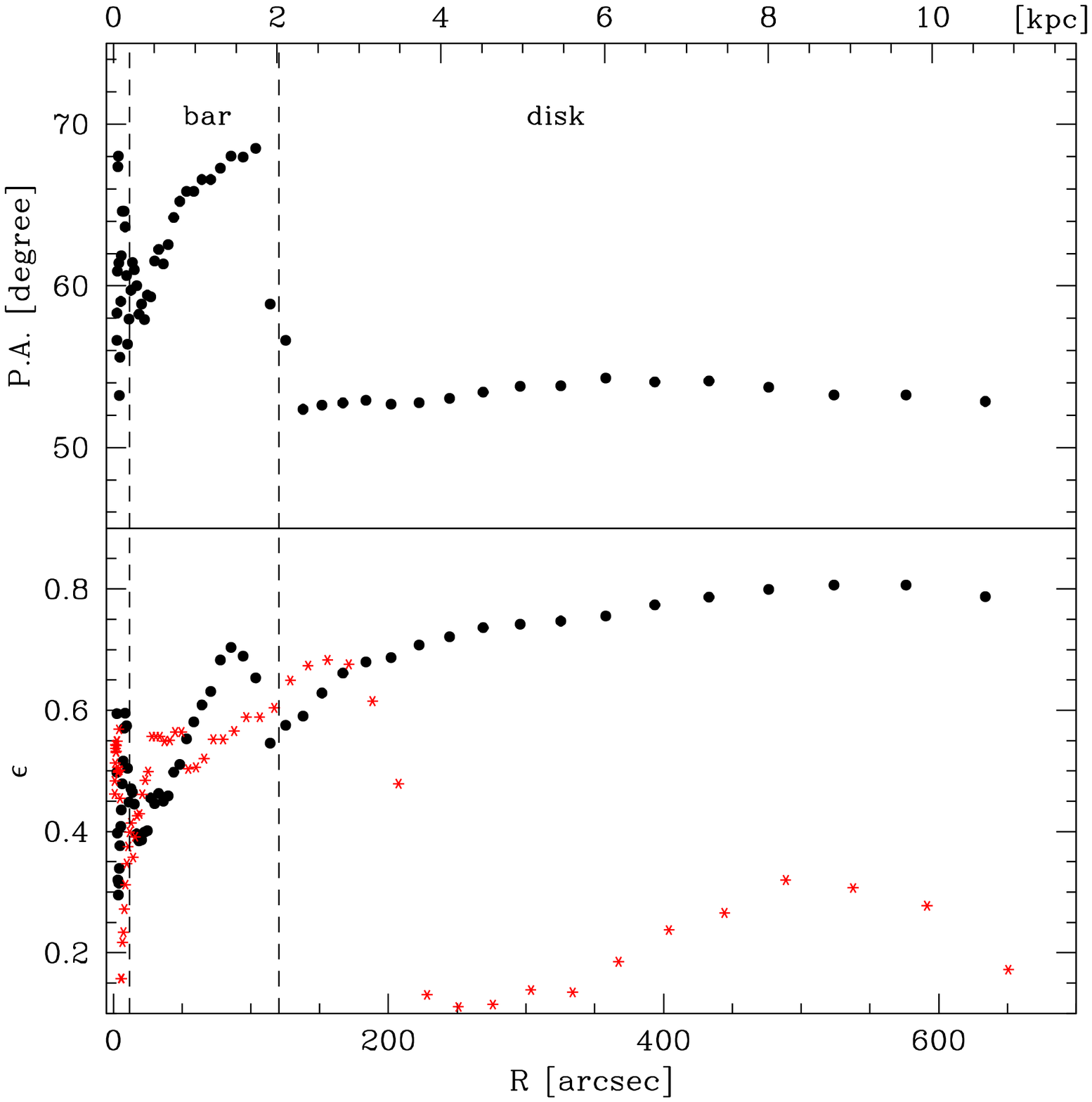}
\includegraphics[width=9cm]{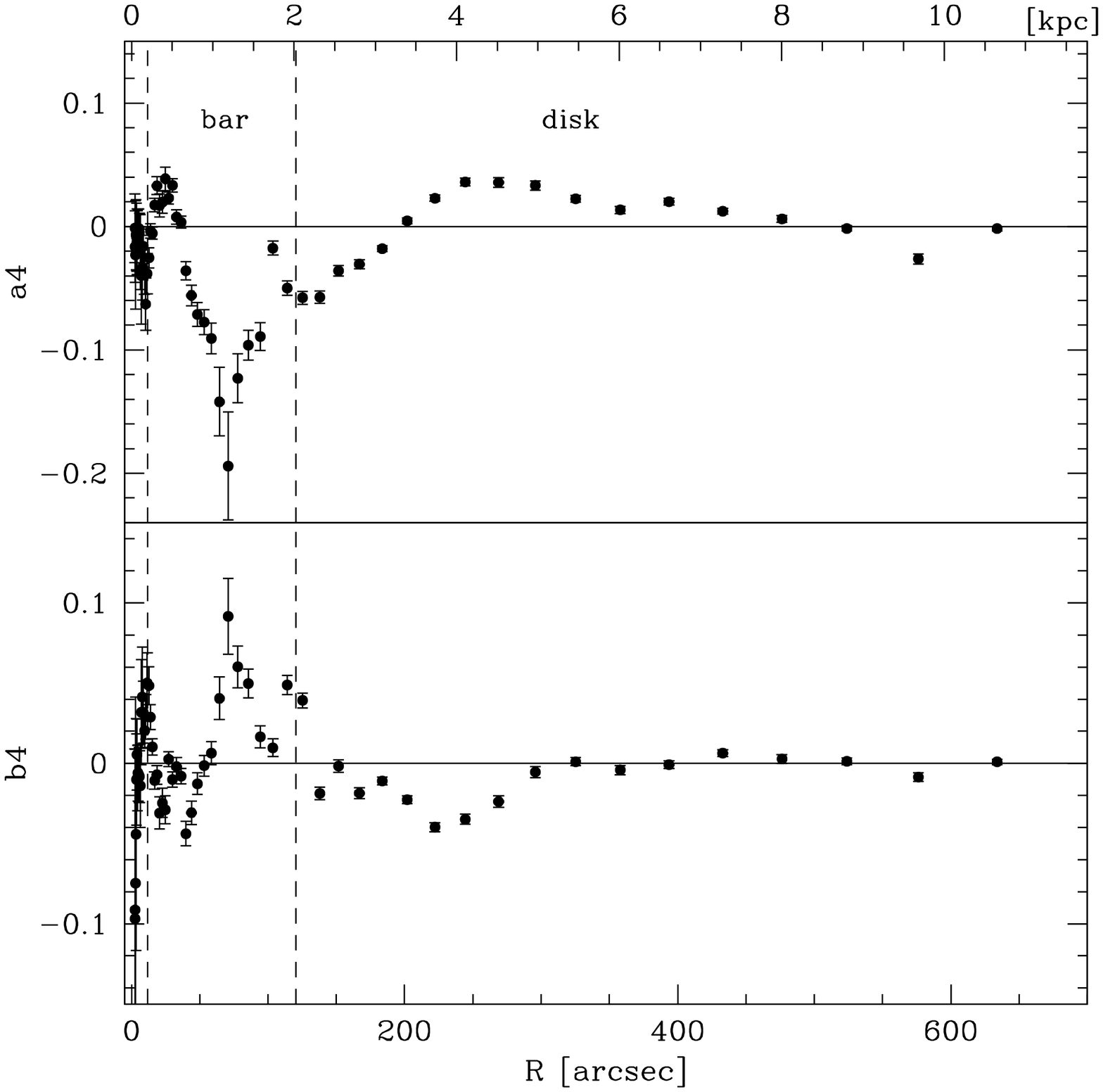}
\caption{Left panel: Average profiles of P.A. (top panel) and
  ellipticity (bottom panel) obtained with ELLIPSE from the Ks-band
  image plotted against the isophote major axis. The red points are
  the measured ellipticity values from the deprojected image (see
  Sec.~\ref{ellfit_dp}). Right panel: $a4$ and $b4$ coefficients of
  the fourth harmonic term in the Fourier series fit to the isophote
  deviations from pure ellipses. The average error on ellipticity is
  $\sim0.002$ and on the P.A. is $\sim 4\deg$. The long-dashed lines
  delimit the regions where the main components of the galaxy
  structure are located. The short-dashed line corresponds to the
  break radius of the disk (see also Sec.~\ref{color}).}
              \label{ellipse}
   \end{figure*}

Position Angle (P.A.) and ellipticity   ($\epsilon = 1-b/a$, where
  $b/a$ is the axial ratio of the ellipses) profiles, in the Ks band
where the dust absorption is minimal, are shown in the left panel of
Fig.~\ref{ellipse}. The deviations of the isophotes from pure ellipses
are estimated by the $a4$ and $b4$ coefficients that are related to the
fourth harmonic term of the Fourier series. The $a4$ and $b4$ profiles as
function of radius are shown in the right panel of Fig.~\ref{ellipse}.
In the regions where the bar dominates the light ($12 \le R \le 120$
arcsec) the ellipticity increases from 0.4 to 0.7 at $R=87.5$~arcsec
($\sim$1.5~kpc), and decreases to about 0.55 at $R=150$~arcsec  
  ($\sim$2.5~kpc).

In these range of radii, a large twisting is observed. The P.A. varies
by about 10~degrees as it moves from $60 \lesssim P.A. \lesssim
70$~degrees, coincident with the peak in ellipticity at
$R=87.5$~arcsec. At larger radii ($R \ge 150$ arcsec), in the disk
region, the P.A. decreases to 52~degrees and both the ellipticity and
P.A. remain almost constant until the outermost measurements, at values 
of 0.8 and $\sim$52~degrees, respectively.   This implies that
  the apparent axial ratio of the disk is $b/a \sim 0.2$ and the disk
  inclination is $74 \pm 3$~degrees. This value is consistent with the
  value of $72 \pm 2$~degrees derived by \citet{P91} from the HI data of
  NGC~253.

The $a4$ and $b4$ profiles (see Fig.~\ref{ellipse}, right panel)
suggest that the isophotes in the inner regions of the bar ($20 \le R
\le 50$ arcsec,  $\sim 0.33 - 1$~kpc) have a boxy shape, with $b4
\sim -0.04$ and $a4 \sim 0.045$, while they become more disk-like at
larger radii ($50 \le R \le 150$ arcsec,   $\sim 0.84 - 2.5$~kpc), where $b4$ reaches a maximum
of $0.06$ and $a4$ a minimum of about $-0.06$ ($70 \le R \le 90$~arcsec,   $\sim 1.2 - 2.7$~kpc).  
Between 150 and 300 arcsec   (i.e., $\sim 1 - 5$~kpc), where the bar connects with the spiral
arms, the isophotes are once more boxy ($b4 \sim -0.04$).  In the disk regions, for $R \ge 300$~arcsec   ($\ge 5$~kpc), the fit of the isophotes is consistent with pure ellipses, i.e. $a4$ and $b4 \simeq 0$.

\begin{figure*}[!ht] 
%\centering
\includegraphics[width=9cm]{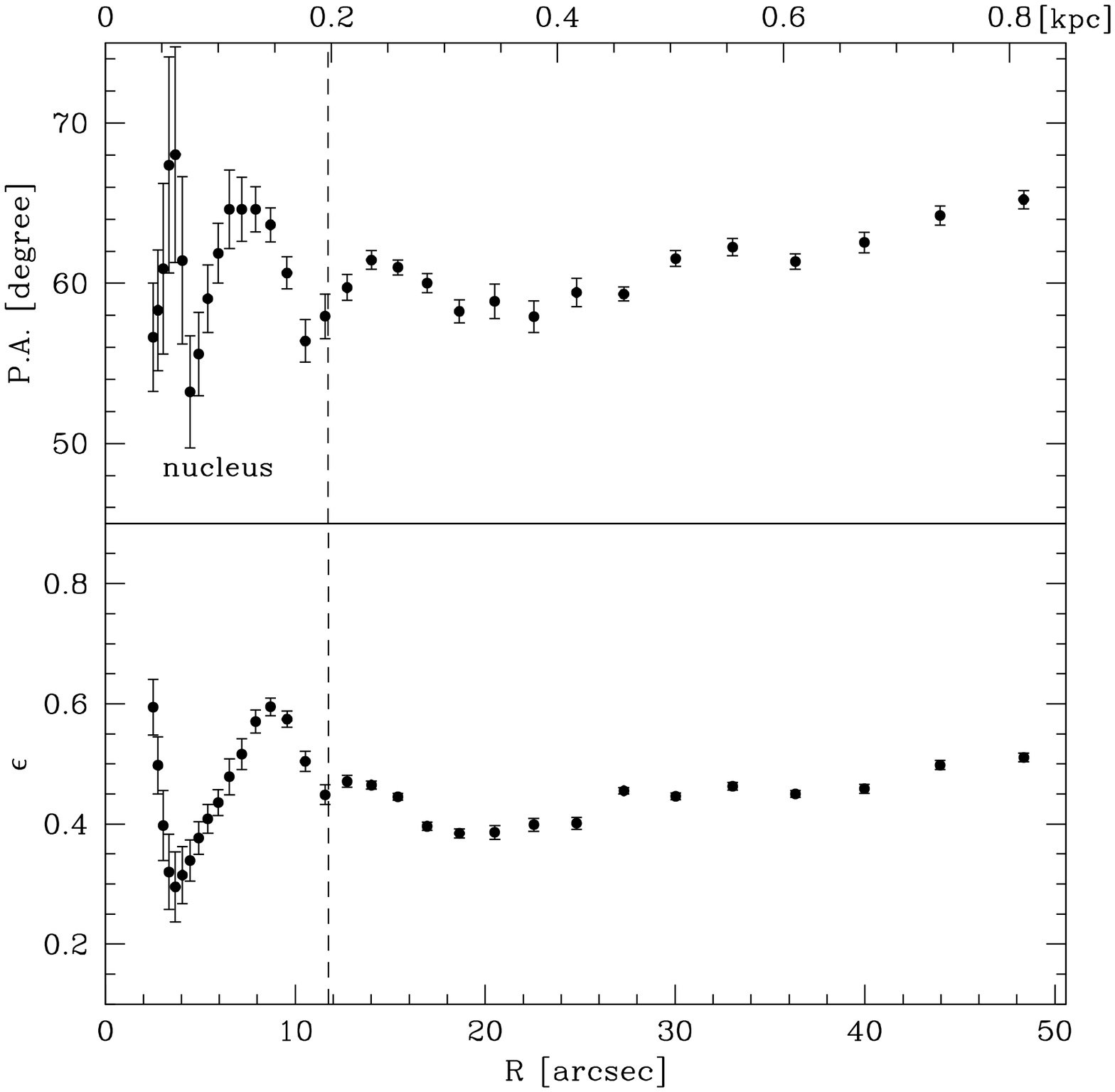}
\includegraphics[width=9cm]{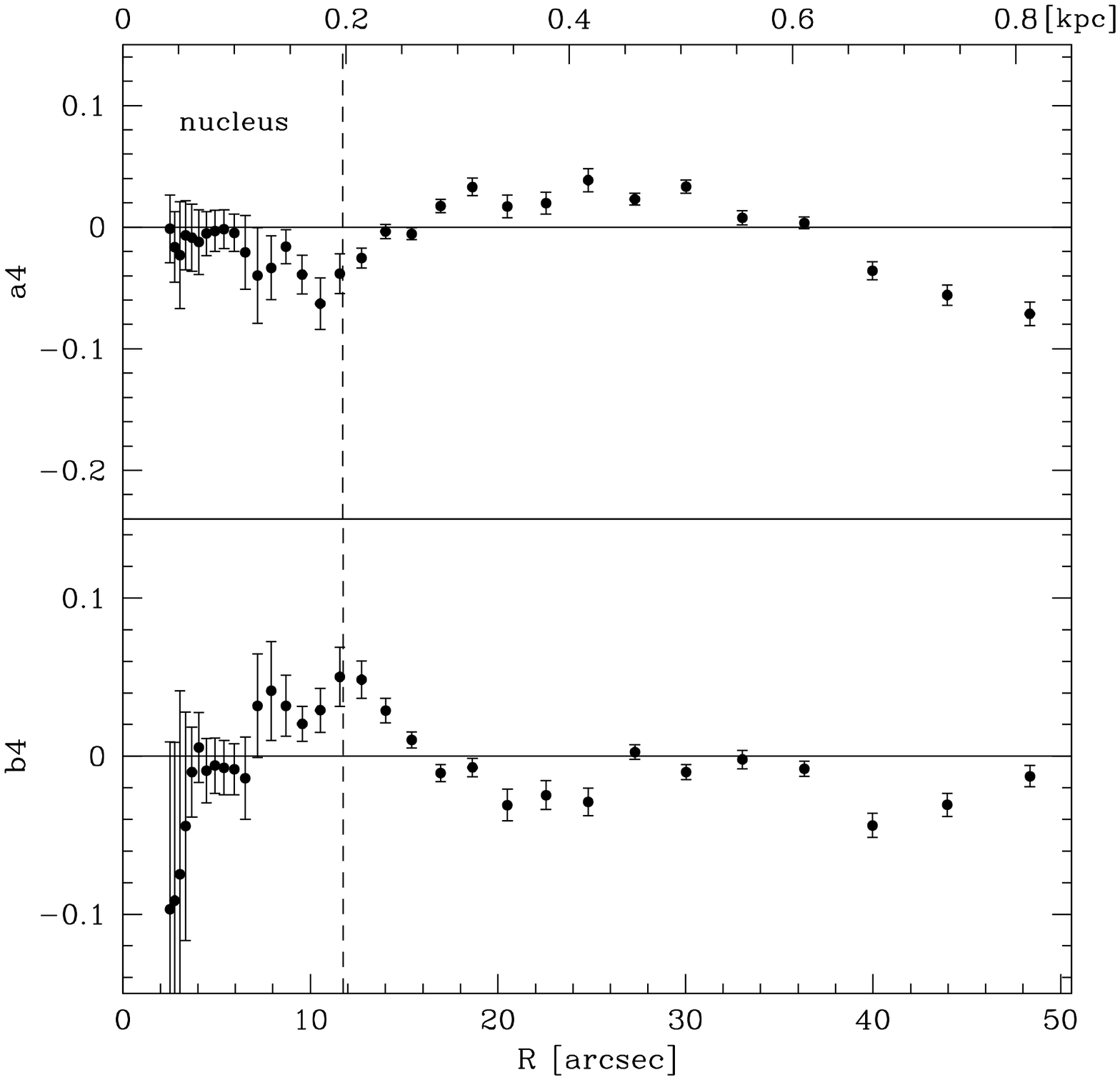}
\caption{Enlargement of the central 50 arcsec region for the P.A. and
  ellipticity (left panels) and for the $a4,\, b4$ profiles (right
  panels). The dashed lines delimit the regions where the main
  components of the galaxy structure are located.}
              \label{ellipse_zoom}
   \end{figure*}

{\it The nuclear region - } In Fig.~\ref{ellipse_zoom} we show an
enlargement on the central 50 arcsec region for the P.A., $\epsilon$ (left
panels), and for the a4, b4 profiles (right panels) derived by the fit
of the isophotes.  Inside $R \le 15$ arcsec   ($\simeq$0.25~kpc), where the  nuclear
  ring is observed (see Fig.~\ref{zoom}), the ellipticity reaches a
maximum of about 0.6 at $R=10$~arcsec ($\sim$0.17~kpc), and subsequently decreases to
about 0.4 at $R=20$~arcsec ($\sim$0.34~kpc). In this region, the P.A. increases from 45
to 58~degrees. This indicates that the nuclear ring is almost as
flat as the bar, but its P.A. differs by about 16 degrees with respect
to the bar P.A.  (as measured from their outer isophotes, i.e. $R=15$~arcsec for the nuclear ring and $R=150$ arcsec for the bar). Inside these regions, the isophotes are more boxy with a maximum
of $b4 \sim 0.03$ at $R \sim 15$~arcsec, ($\sim$0.25~kpc), (see Fig.~\ref{ellipse_zoom}, right panels).

 {\it The outer disk - } In the region of the outer disk, both J and Ks profiles show abrupt changes of slope for $R \ge 550$~arcsec  ($\sim$9.2~kpc). As shown in the right panel of  Fig.~\ref{fit_log}, the surface brightness profile has a sharp
  decline with respect to the "inner" regions of the disk, producing a  {\it downbending Type~II profile}, according to the classification  of light profiles in disk galaxies \citep[see e.g.][]{PhT06, Erw08}. This feature can be reasonably ascribed to the effect of  disk truncation, as shall be discussed in Sec.~\ref{disk}.

In order to derive the break radius $R_{br}$ of the disk in NGC~253,
we performed a least-square fit of the azimuthally averaged surface
brightness profiles, restricted to the region where the contribution
to the light of the bulge plus bar is negligible, i.e. for $R\ge 120$~arcsec,   ($\ge$2~kpc), (see Fig.~\ref{fit_log}, right panel). Two exponential functions are used to describe the {\it inner} and {\it outer} disk:
\begin{equation}
\mu^{in,out}(R)= \mu^{in,out}_{0} + 1.086 \times R/r^{in,out}_{h}
\end{equation}
where $R$ is the galactocentric distance, $\mu^{in,out}_{0}$ and $r^{in,out}_{h}$ are the central surface brightness and scale length of each of the two exponential components. In the Ks band, the best fit to the structural parameters 
are summarized in Table~\ref{tabgalfit}.
%In the Ks band, the best fit gives the following values for the above structural parameters, that characterises the inner and outer disk: $\mu^{in}_0=15.58 \pm 0.05$~mag/arcsec$^2$ and $r^{in}_h = 173 \pm 4$~arcsec for the inner disk, and $\mu^{out}_0=12.56 \pm 0.07$~mag/arcsec$^2$ and $r^{out}_h = 92.6 \pm 0.5$~arcsec for the outer disk. A summary of the fitted parameters is given in Table~\ref{tabgalfit}. 
For the J band profiles, we found $\mu^{in}_0=16.67 \pm 0.05$~mag/arcsec$^2$ and $r^{in}_h = 193
\pm 5$~arcsec   ($\sim$3.2~kpc) for the inner disk, and $\mu^{out}_0=16.00 \pm 0.07$~mag/arcsec$^2$ and
$r^{out}_h = 159 \pm 1$~arcsec   ($\sim$2.7~kpc) for the outer disk. 

The inner disk scale length is larger than the outer one, being
$r^{in}_h/r^{out}_h = 1.87 \pm 0.02$ in the Ks band, and
$r^{in}_h/r^{out}_h = 1.22 \pm 0.03$ in the J band.  This is also observed for other galaxies with Type~II profiles \citep{Kim2014}. 
The break radius $R_{br}$ correspond to the radius at which the magnitudes of the two exponential laws are the same, within $2\sigma$. The best fits are shown in the right panel of Fig.~\ref{fit_log}, and the values of
$R_{br}$ are: $R^J_{br} = 553.20 \pm 0.01$~arcsec and $R^{Ks}_{br} = 554.05 \pm 0.01$~arcsec  (at $\sim9.3$~kpc). 

\subsection{Deprojected image of NGC~253 }\label{ellfit_dp}

\begin{figure}[!ht] 
%\centering
\includegraphics[width=9cm]{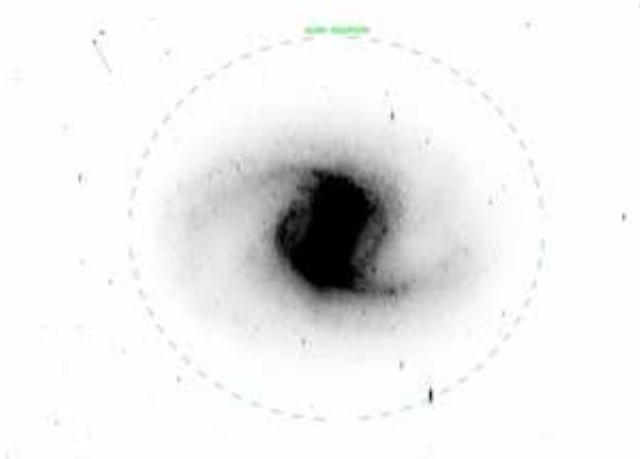}
\caption{Deprojected image of NGC~253 in Ks band.   The image size
    is $39.5 \times 26.6$~arcmin. The dashed green line indicates the
    outer isophote, which is almost circular.}
              \label{imaK_dp}
   \end{figure}

We computed the deprojected image of NGC~253, both in the J and Ks
bands to measure the bar structure. These measurements are discussed
in detail in Sec.~\ref{bar}. According to Gadotti et al. (2007), an
efficient way to deproject an image is to stretch the image in the
direction of the line of nodes by using the IRAF task IMLINTRAN with
the constraint that flux is conserved.   For NGC~253, we adopted the
  P.A. of the outer disk (i.e. P.A.= 52 degrees) as the P.A. of the
  line of nodes and the inclination angle of $i=74$~degrees (see
  Sec.~\ref{ellfit}). The deprojected image of NGC~253 in the Ks-band
  is shown in Fig. \ref{imaK_dp}. This is very similar to that derived
  by \citet{Dav10}.

In order to derive the deprojected ellipticity for the bar, an
important parameter to constrain the bar strength and bar length (see
Sec.\ref{bar} for a detailed discussion), we applied   the isophotal
  analysis to the deprojected image in the Ks-band. The
deprojected ellipticity radial profile is shown in Fig.~\ref{ellipse}
(middle-left panel, red points). In the bar region, the pattern is
very similar to that of the ellipticity radial profile for the
observed image, although, as expected, the ellipticity value and the
position of the peak change. The deprojected bar appears longer and
less eccentric. The ellipticity reaches its maximum of about 0.65 at
larger distances from the center, i.e. R=160~arcsec   ($\sim$2.7~kpc). As expected, the
outer disk becomes more circular once deprojected.

\subsection{Light and Dust distribution}\label{color} 

To describe the light distribution of the major structural components
of NGC~253 we extracted the one dimensional (1D) light profiles along
the disk major axis (P.A. = 52 degrees) and along the bar major axis
(P.A. = 72 degrees) in the J and Ks-band shallow images.   These are derived by
  averaging the counts in a wedge with an opening angle of 5~degrees that extends out to
  the outer limiting radius derived in Sec.~\ref{2mass}.
Furthermore, we produced the 2D J-Ks color map and extract the J-Ks
color profiles along P.A. 52 and 72 degrees. Even if the light
profiles in the J band are much more affected by dust, the same
components are detected in both J and Ks-band images.

\begin{figure*}[!ht] 
\centering
\includegraphics[width=9cm]{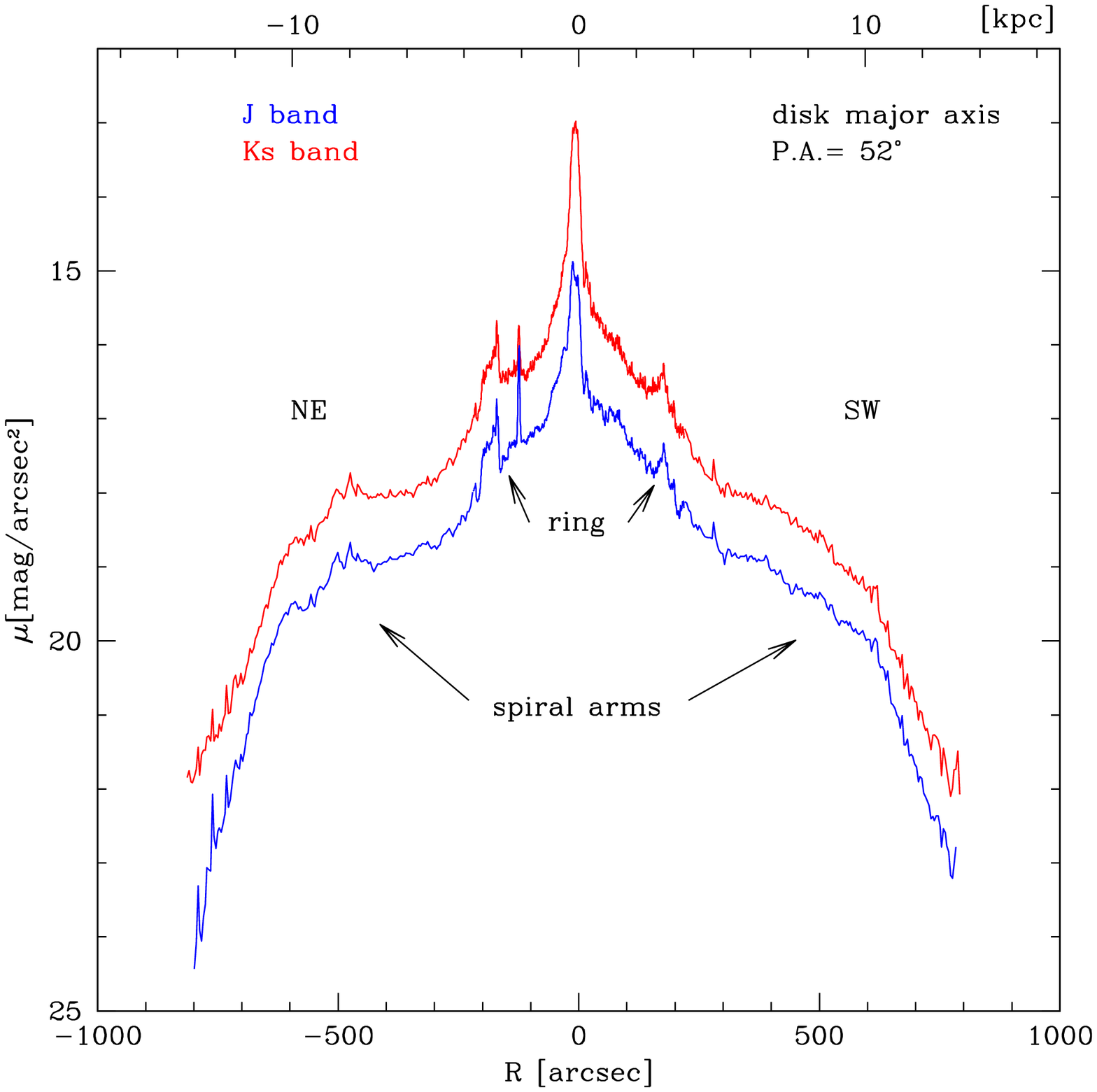}
\includegraphics[width=9cm]{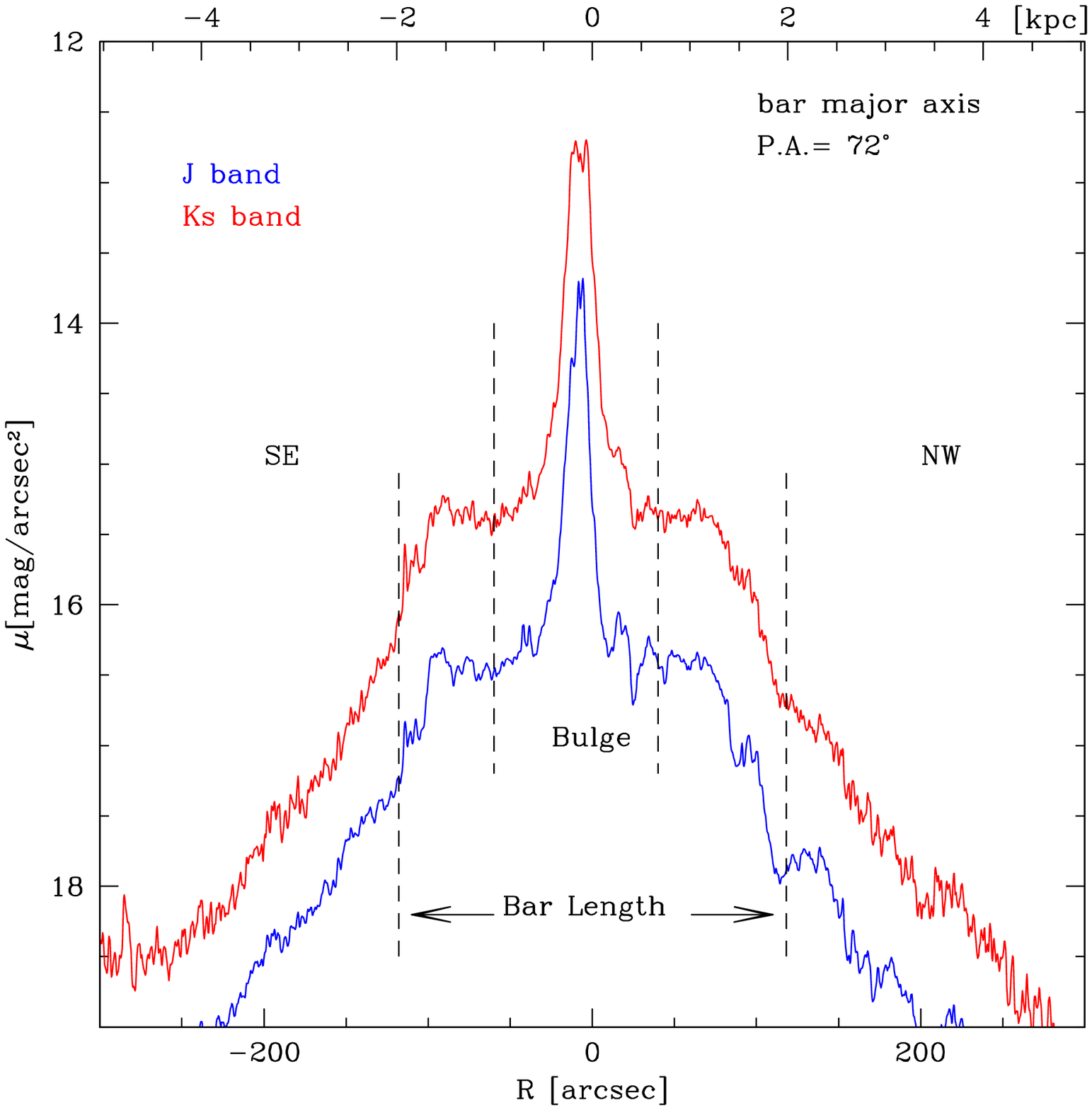}
\caption{Surface brightness profiles in the J and Ks-bands extracted
  along the disk major axis ($P.A. = 52$ degrees, left panel) and
  along the bar major axis ($P.A. = 72$ degrees, right panel).}
              \label{profK}
   \end{figure*}

{\it Disk structure: bar, ring and spiral arms -} The J and Ks surface
brightness profiles along the disk major axis and along bar major axis
(Fig.~\ref{profK} left and right panel, respectively) are
characterized by a bright and rather concentrated bulge. Over the
average light distribution, some peaks are observed. On the NE side,
two peaks are detected for $100 \le R \le 200$~arcsec   ($1.7
  \lesssim R \lesssim 3.4$~kpc); on the SW side another peak is
located at $R \sim 158$~arcsec ($\sim$2.6~kpc). They are related to
the ring-like component   enclosing the bar (see Sec.~\ref{morph}
and Fig.~\ref{tileKs}). At larger radii in the disk regions, as
  already detected in the average surface brightness profile derived
  from the isophote fit (see Sec.~\ref{ellfit}), the surface
  brightness has the typical behavior of a down-bending Type-II
  profile, observed for disk galaxies \citep{PhT06, Erw08}, with a
  break radius of $R_{br} \simeq 553$~arcsec ($\sim 9.3$~kpc). 

The signature of the bar in the light profiles along $P.A. = 72$~degrees (Fig.~\ref{profK} right panel) appears on both sides,
between $60 \le R \le 120$~arcsec  ($\sim 1 - 2$~kpc), as a plateau followed by a
steep decline compared to the underlying exponential disk \citep{Ph00}. This is most evident in the Ks-band.  The distance from the galaxy center at which the turn-over in profile slope occurs (at
about 118~arcsec, or $\sim$2~kpc), can be considered as a rough
estimate of the bar projected length \citep{Lut00}. An estimate of the
bulge length $R_{B}$ is given by the region where light peaks over to
the bar and the exponential disk. We derived $R_{B} \sim 60$ arcsec ($\sim$1~kpc).

\begin{figure*}[!ht] 
\centering
\includegraphics[width=9cm]{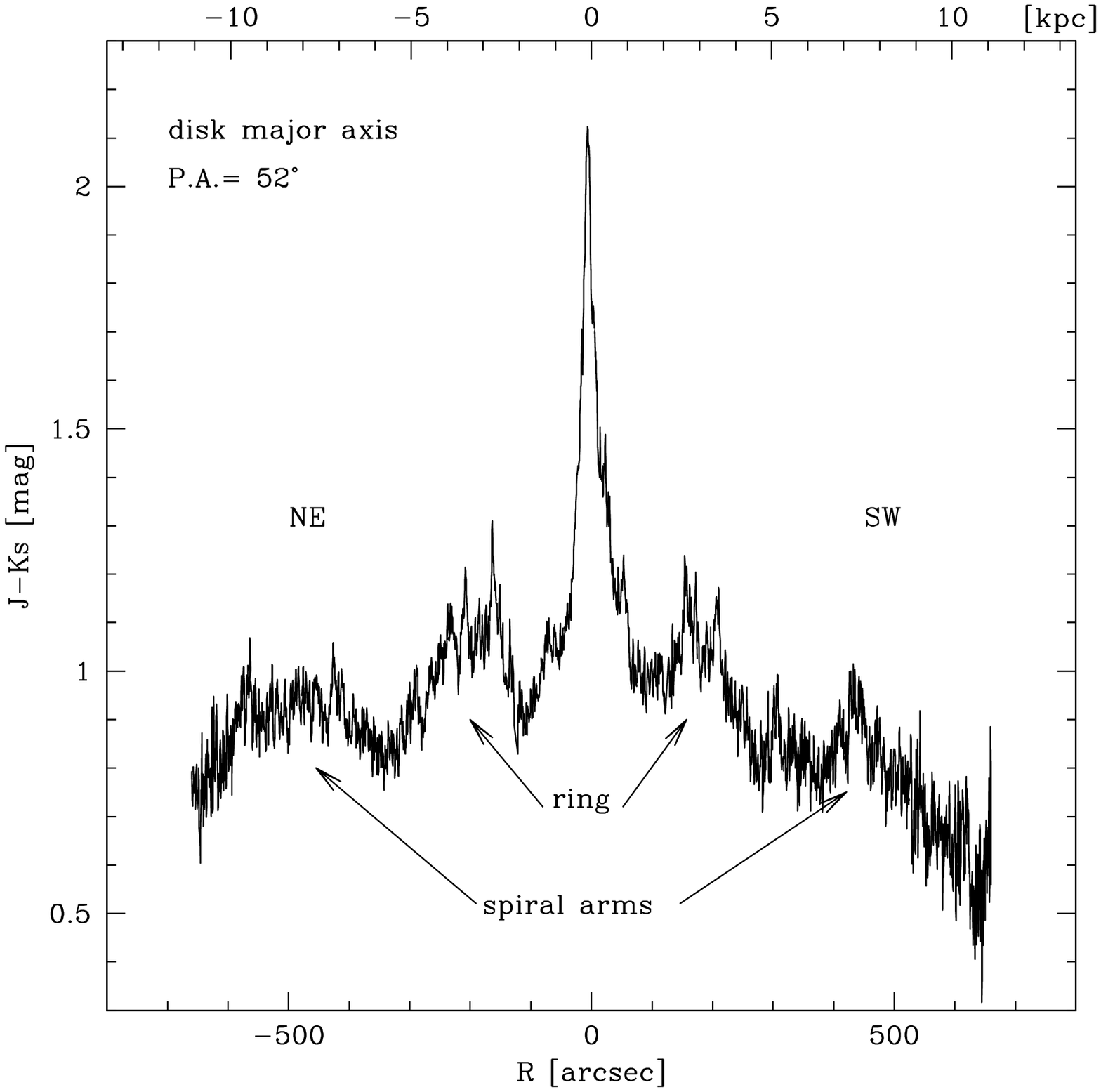}
\includegraphics[width=9cm]{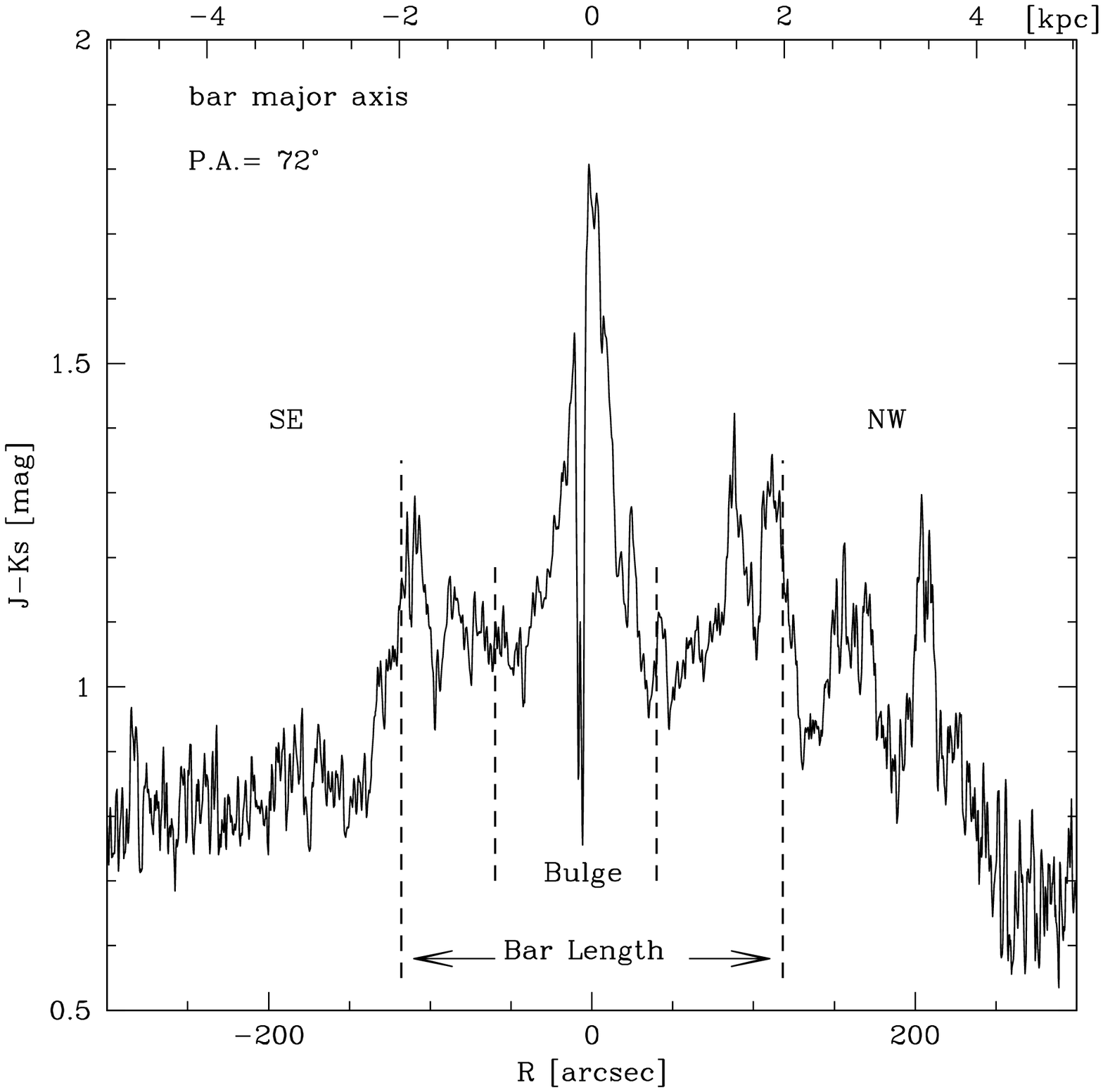}
\caption{J-Ks color profiles extracted along the disk major axis
  ($P.A. = 52$ degrees, left panel) and along the bar major axis
  ($P.A. = 72$ degrees, right panel).}
              \label{profJ_K}
\end{figure*}

Fig.~\ref{profJ_K} shows the J-Ks color profiles along the major axes of the disk (left panel) and bar (right panel). On average, both profiles show that: {\it i)}   at small radii, close to the center,  the galaxy is redder ($J-Ks \sim 2.2$~mag along the major axis) with respect to the outer regions; {\it ii)} the color profiles decrease
rapidly and reach a value of $J-Ks \sim 1$~mag at $R = 50$~arcsec   ($\sim$0.84~kpc) as we
move away from the center; and {\it iii)} the color gradient becomes almost linear with $1 \le J-Ks \le 0.5$~mag to the outermost radii. On both E and W sides along the disk major axis, we associate the peaks
in the surface brightness profiles of Fig.~\ref{profK} (left panel) in
the range $100 \le R \le 300$~arcsec   ($\sim 1.7 - 5$~kpc) with the ring and spiral
arms. Projected on the disk major axis (see Fig.~\ref{profJ_K} - left panel), we see that these features are redder than the underlying disk, with $J-Ks \sim$1.2~mag for the ring and $J-Ks \sim 0.8$~mag for the
spiral arms. At larger radii, in particular for $R \ge R_{br}$,
the disk tends to be bluer as $J-Ks$ varies from 0.9 to about 0.7~mag on
NE side, and from 0.7 to about 0.5~mag  on SW side.  
Between   $50 \le R \le 120$~arcsec  ($\sim 0.84 - 2$~kpc), the color profile along the bar major axis is on
average redder $1.0 \le J-Ks \le 1.3$~mag than the disk. We also report
the presence of two redder ($J-Ks \sim 1.2$~mag) peaks, which correspond
to the bright edges of the bar (see Sec.~\ref{bar} for an detailed discussion).

\begin{figure*}[!ht]  
\centering
\includegraphics[width=16cm]{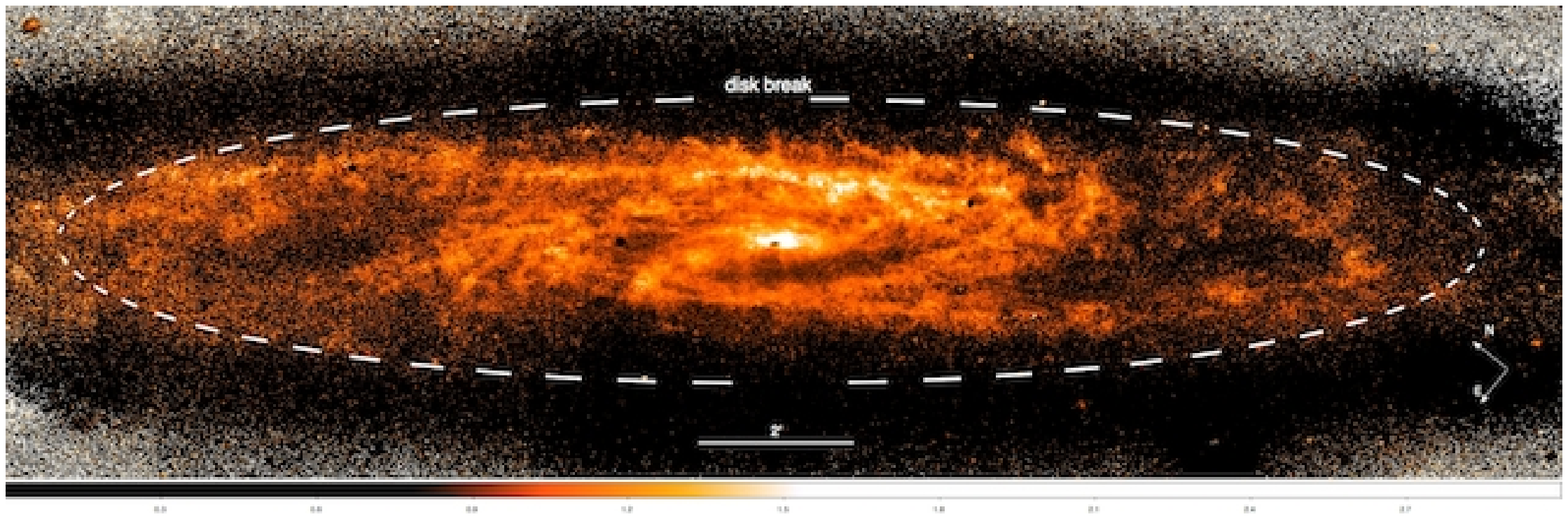}
\includegraphics[width=16cm]{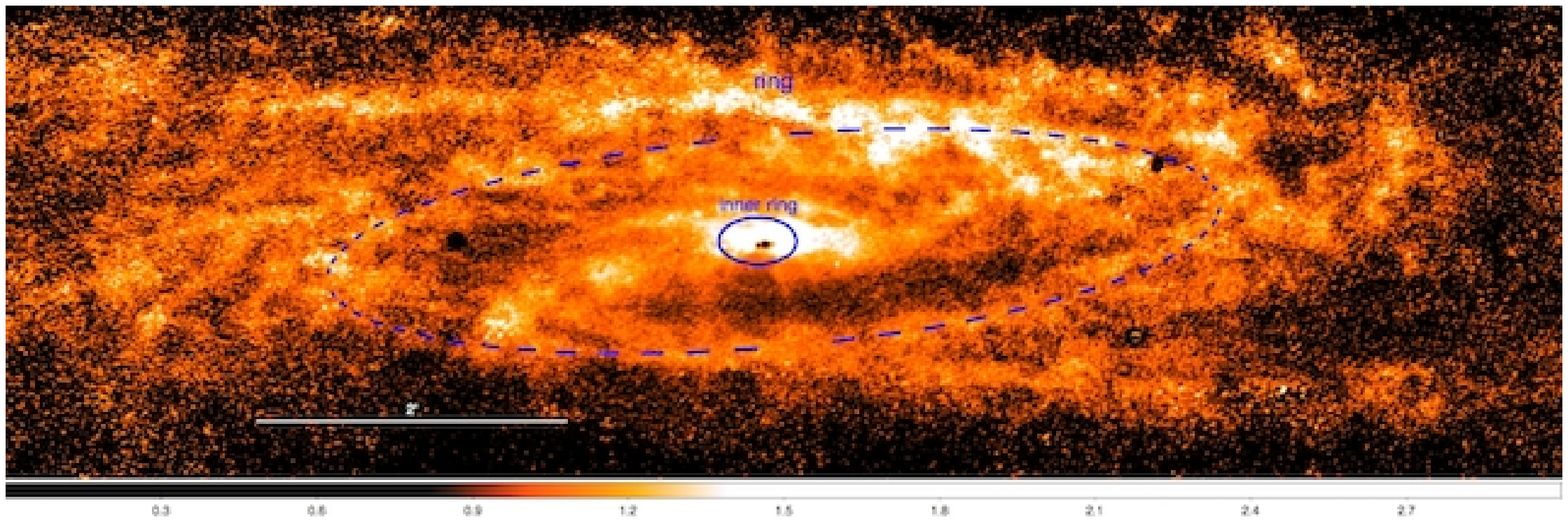}
\caption{J-Ks color map of NGC~253 and an enlargement of the nuclear
  region (bottom panel). Lighter colors are redder J-Ks values.  
    The isophote corresponding to the disk break (white dashed line)
    is overlaid on the image in the top panel. The blue ellipses
    overlaid on the enlarged image (bottom panel) indicate the regions
    where the nuclear rings (dashed line) and the ring at the
    end of the bar (solid line)reside.} 
              \label{colormap}
   \end{figure*}

The two-dimensional J-Ks color distribution is shown in Fig.~\ref{colormap}.  The reddest regions, i.e. $(J-Ks) \ge 1.5$~mag, correspond to the NW arm, 
the edges of the bar along the EW  direction, and the nuclear regions. From the J-Ks color map, one can
  infer that the dust is confined to the spiral arms and the ring in NGC~253.  Dust lanes are curved with
their concave side toward the bar major axis and they curl around the
nucleus at small radii.

\begin{figure*}[!ht] 
\centering
\includegraphics[width=9cm]{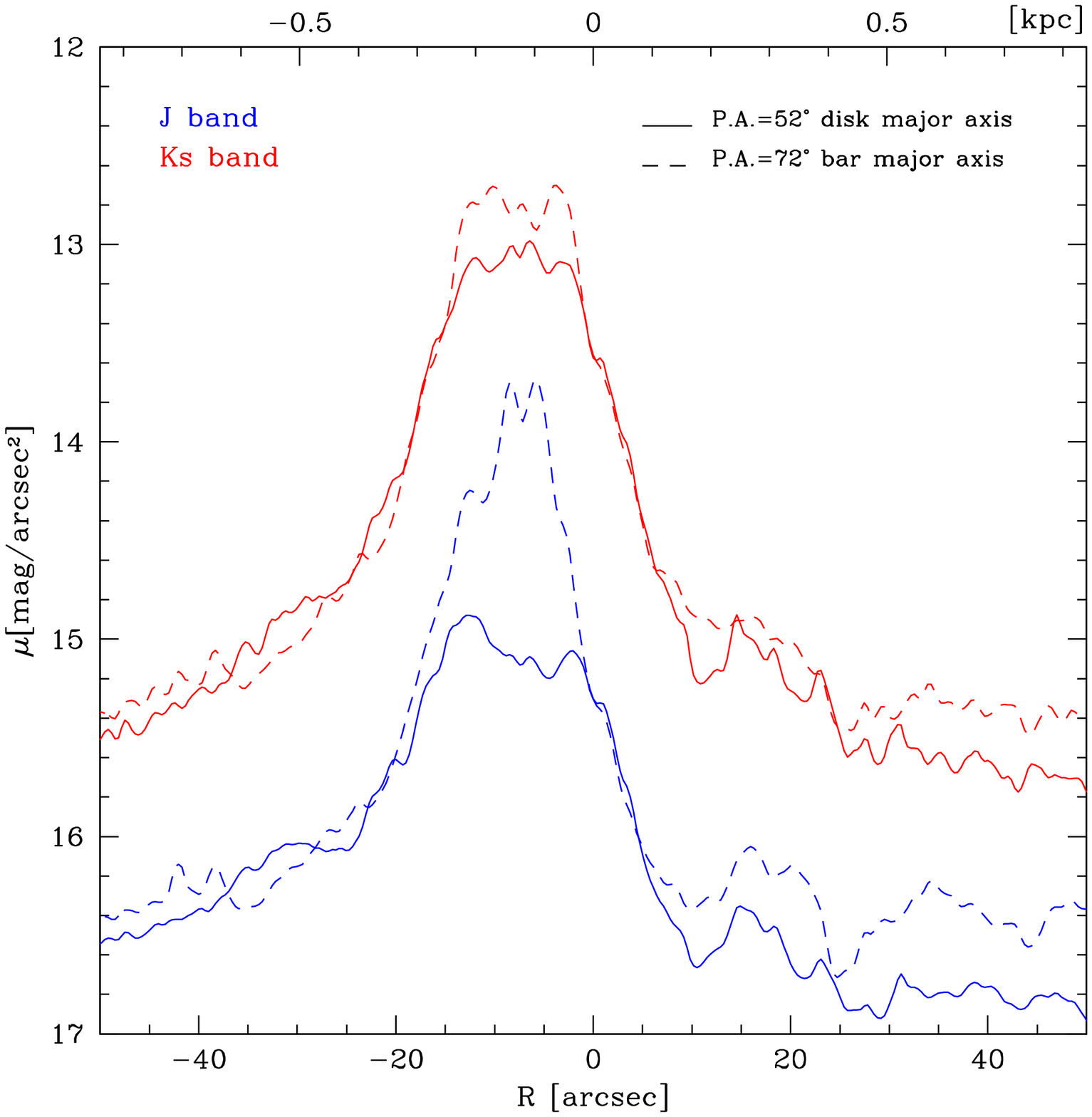}
\includegraphics[width=9cm]{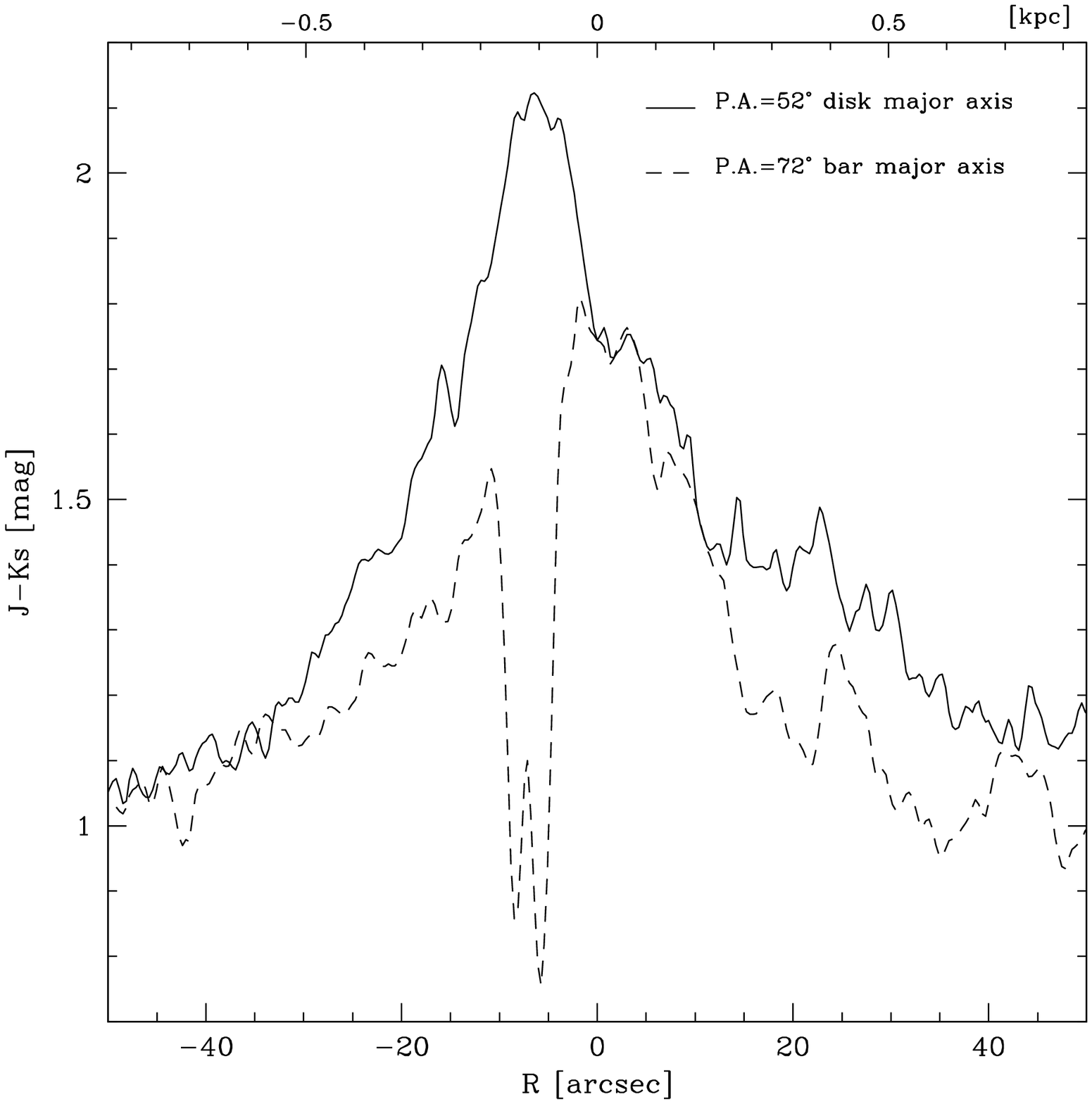}
\caption{Enlargement of the nuclear region. Left panel: Ks (red line)
  and J (blue) surface brightness profiles along the disk (continuous
  line) and bar (dashed line) major axis. Right panel: J-Ks color
  profile.}
              \label{profK_zoom}
   \end{figure*}

{\it The nuclear region - } Fig.~\ref{profK_zoom} shows an enlarged view of both light and color profiles within the nuclear region at $R\le 50$ arcsec ($\le$0.84~kpc).  Inside 15~arcsec  ($\le$0.25~kpc), the light profiles along the disk
and bar major axes are very irregular and asymmetric with respect to the center\footnote{The central pixel adopted to extract the light  and color profiles coincides with the kinematic center found by
  \citet{MS10}, given at the beginning of the Sec.~\ref{phot}.}, and
reflect the peculiar nuclear structure of NGC~253. Several small peaks
of light are observed and for $R \le 10 $~arcsec, the surface
brightness remains almost constant at about $\mu_J \sim 15.2$~mag~arcsec$^{-2}$ and $\mu_{Ks} \sim 13$~mag~arcsec$^{-2}$ (see Fig.~\ref{profK_zoom}, left panel). Along the bar major axis, the
light profiles are much more peaked and reach a maximum value of
$\mu_{Ks} \sim 13.8$~mag~arcsec$^{-2}$. Several compact sources,
brighter than the underlying diffuse light distribution in the  
  nuclear ring (see Sec.~\ref{morph} and Fig.~\ref{zoom}), can be
identified as being responsible for the observed peaks of the surface
brightness profiles. The flat regions can be due to dust absorption
that is still quite high even in the Ks-band. This is suggested by the
J-Ks color profile which increases by more than one magnitude within
20~arcsec from the galaxy center (see Fig.~\ref{profK_zoom}, right
panel). Inside $R \le 5$ arcsec   ($\le$0.08~kpc), the J-Ks color profile along the bar
major axis shows a rapid decrease toward bluer colors, where the
minimum is $(J-Ks) \sim 0.2$~mag. This corresponds to a ``blue hole'' in the
J-Ks color map (see the bottom panel of Fig.~\ref{colormap}).  
   The color map also shows redder filamentary structures, with $(J-Ks)  \sim 1$~mag, that are most evident in the SE regions of the galaxy,  starting from the galaxy nucleus and extending in the orthogonal
  direction. Such ``polar'' filaments were observed by \citet{SW92}, and
  recently confirmed by ALMA observations \citep{ALMA}; they could be
  made of dust associated with the outflow material from the burst
  of  newly formed stars. Similar features are observed in the
  edge-on spiral galaxy NGC~891 \citep[see][and references therein]{Wh09}.

\section{Two dimensional model of the light distribution in the Ks band}\label{galfit}

The current morphological analysis suggests that NGC~253 is dominated
by three main structures at different distances from the galaxy
center: the bright nucleus, the bar and the extended disk. In addition
to the smooth light distribution of these main components we observe
the following substructures (see Sec.~\ref{morph} and Sec.~\ref{phot},   and Fig.~\ref{tileKs}, Fig.~\ref{FM} and Fig.~\ref{zoom}): a nuclear ring, a ring  at the end of the bar and spiral arms.  In order to measure the structural
parameters of the main galaxy components and quantify the observed
substructures, in terms of their total luminosity and radial
extensions, we adopted the following approach. In the Ks-band, which is
less perturbed by dust, we performed {\it i)} a 1D least-square fit of
the light profiles along disk and bar major axes with masked
substructures; {\it ii)} produced a "maximum" symmetric two
dimensional (2D) model by assuming that the galaxy light distribution
arises solely from the above three components (i.e. bulge, bar and
disk); and {\it iii)} subtracted the 2D model from the original image
to create a residual image with only the substructures. We then
measureed the luminosity and extensions of the substructures on this
residual image.

\begin{figure*}[!ht] 
\centering
\includegraphics[width=9cm]{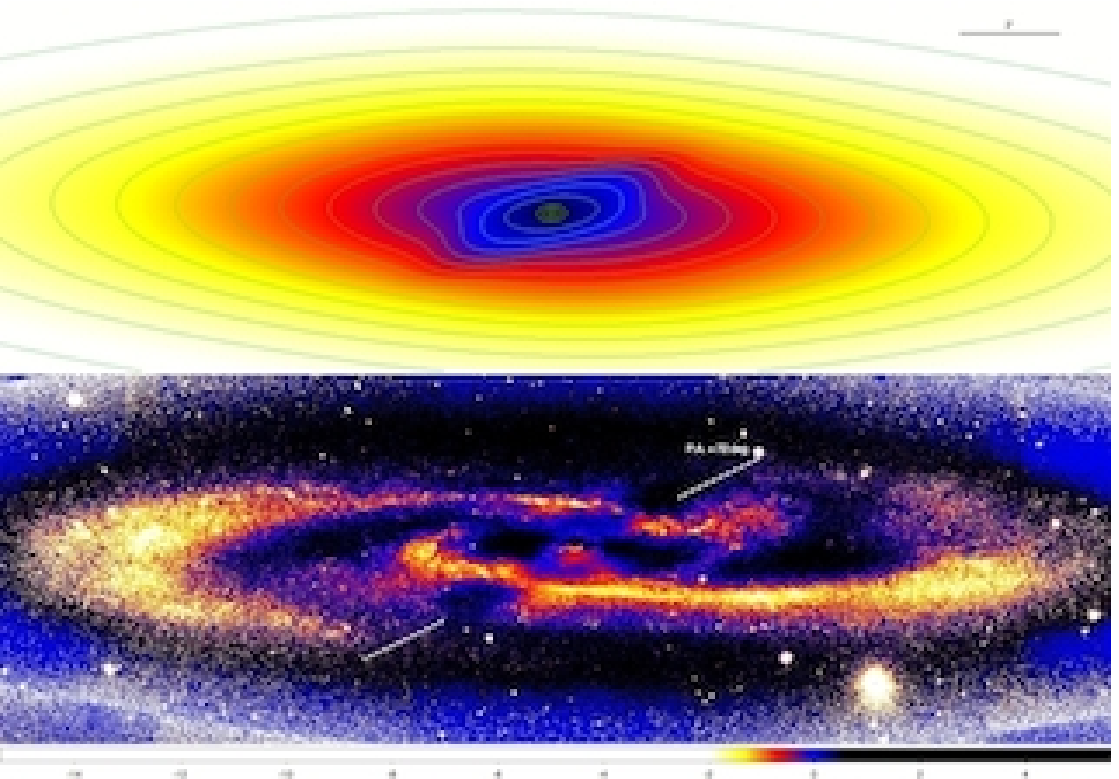}
\includegraphics[width=9cm]{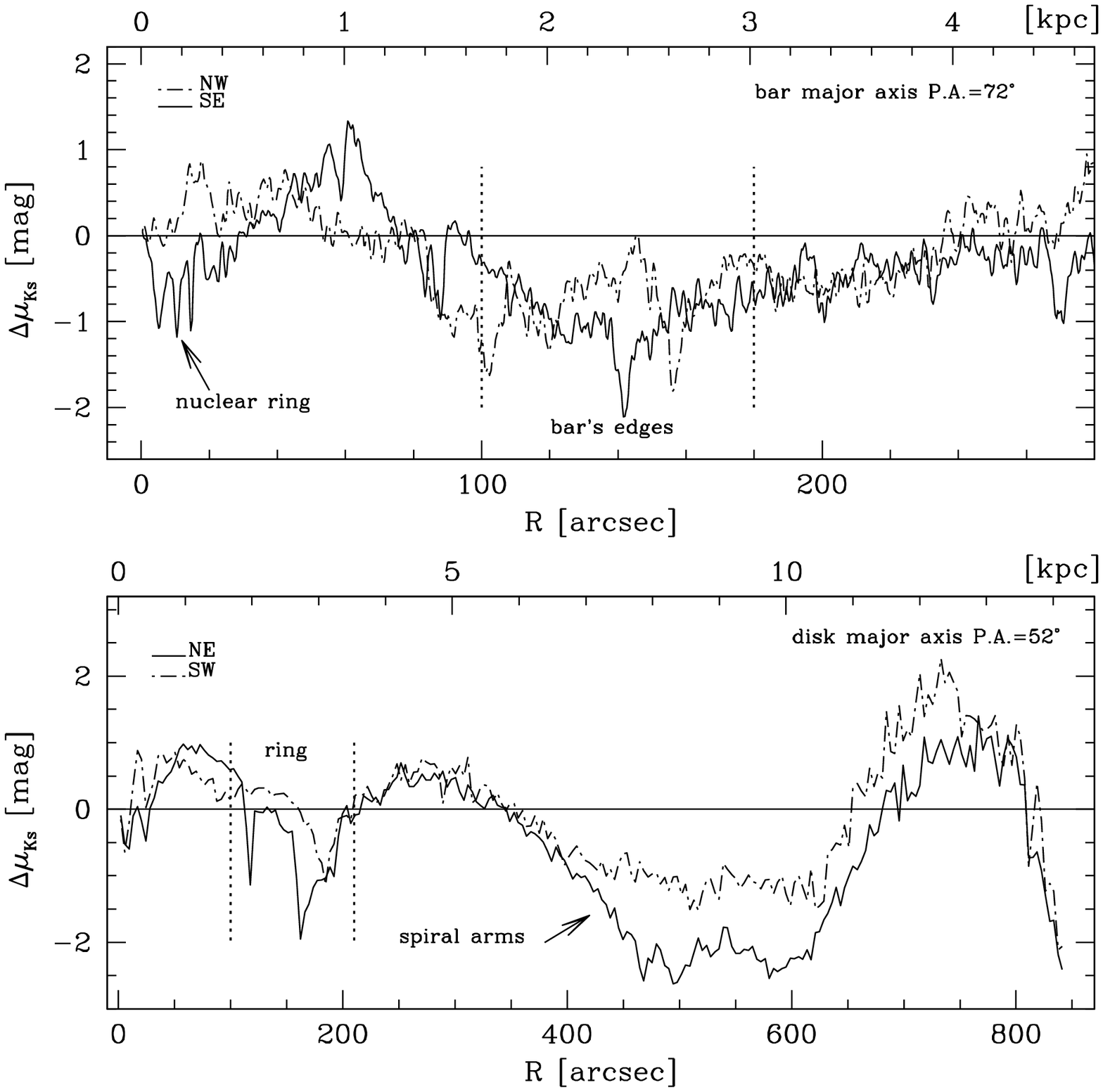}
\caption{Top-left panel: Galfit 2D model computed using two Sersic
  laws for bulge and bar, and a double exponential law for the
  disk. Bottom-left panel: residual image derived by subtracting the
  2D model from the galaxy image, in a logarithmic scale.  Right
  panels: differences between the observed and fitted light profiles
  along the disk major axis (bottom panel) and along the bar major
  axis (top panel).  $\Delta \mu < 0$ where the galaxy is brighter
  than the model.}
              \label{2dmod}
   \end{figure*}

  The light distribution of the bulge and bar in NGC~253 is modeled
  by a Sersic law \citep{Sersic, Kim2014}
\begin{equation}
\mu (R) = \mu_e + k(n) \left[ \left( \frac{R}{r_e}\right)^{1/n} -1\right]
\end{equation}
where $R$ is the galactocentric distance, $r_e$ and $\mu_e$ are the
effective radius and effective surface brightness, and $k(n)=2.17 n - 0.355$. For the disk, as
described in Sec.~\ref{ellfit}, we adopted a double exponential law to
describe the down-bending profile of this component (see also
Fig.~\ref{fit_log}). The maximum symmetric 2D model is made by using
the GALFIT package \citep{Peng02}, where the scale radii ($r_e$ and
$r_h$) and the shape parameters $n$ are fixed to those values of the
1D fit, while the total magnitudes, axial ratios and P.A.s are left
free. A summary of the structural parameters for each component is
listed in Table~\ref{tabgalfit} and the results are shown in
Fig.~\ref{2dmod}. In the 2D residual image (bottom-left panel of
Fig.~\ref{2dmod}) all the substructures characterising NGC~253 are
clearly visible as regions where the galaxy is brighter than the
model: the nuclear ring, the bright regions at the end of the bar from
which the two prominent spiral arms emerge, and the ring which
encloses the bar.

The difference between the observed and fitted light profiles, along
the disk major axis and along the bar, is shown in the right panels of
Fig.~\ref{2dmod}. Along the disk major axis (bottom panel), the
largest deviations ($\Delta \mu \ge 1$~mag) are in the regions
corresponding to the ring at $150\le R \le 200$~arcsec ($\sim$3~kpc),
where the bright bumps are $\sim1-2$~mag brighter than the underlying
average light distribution. For $300 \le R \le 700$~arcsec   ($\sim 5 - 11.8$~kpc), the
extended bumps,   $\sim 1.5 - 2.5$~mag brighter than the
double-exponential fit, are due to the spiral arms. The bulge
is less prominent with respect to the other components in NGC~253,
since the Bulge-to-Total ratio $B/T = 0.043$.

Along the bar's major axis (Fig.~\ref{2dmod}, upper panel) the fit is very good (better
than 0.2~mag), except for the nuclear regions $R \le 20$~arcsec
($\sim$0.34~kpc), where the peaks in the galaxy brighter than the
model are associated with the nuclear ring (see Sec.~\ref{morph} and Fig.~\ref{zoom}), and
for $100 \le R \le 180$~arcsec ($\sim$2.3~kpc) where the largest
deviations of about 2~mag occur close to the edges of the bar.

\begin{table*}
\caption{Structural parameters for the 2D model of the light  distribution of NGC~253 in the Ks band}
  \label{tabgalfit} 
  \centering
\begin{tabular}{c c c c c c c c c c}     % 8 columns 
\hline\hline       
Component & Model & $m_{tot}$ & $\mu_{e}$ & $\mu_{0}$ & $r_e$ & $r_e$ & $r_h$ & $r_h$ & n \\
  &  & mag & mag arcsec$^{-2}$ & mag arcsec$^{-2}$ & arcsec  & kpc & arcsec & kpc &  \\  
\hline                    
Bulge         & Sersic  &  7.31  & $15.49 \pm 0.05$ &                             & $9.1 \pm 0.5$ & 0.15 &   &  & $0.76 \pm 0.01$ \\ 
Bar             & Sersic  & 6.45  & $19.38 \pm 0.05$ &                              & $81 \pm 1$     & 1.4   &   &  &  $0.30 \pm 0.01$ \\ 
Inner Disk   & exp      & 4.45  &                             & $15.58 \pm 0.05$ &                        &          & $173 \pm 4$ & 2.9 &  \\ 
Outer Disk  & exp      & 5.19  &                             & $12.56 \pm 0.07$ &                        &          & $92.6 \pm 0.5$ & 1.5  &  \\ 
\hline                  
\end{tabular}
\tablefoot{  The first column indicates the different   structures in NGC~253. In Col.2 we indicate the empirical law
    adopted to fit the light distribution. In Col. 3 we list the total magnitude corresponding to each component; and from Col.4 to Col. 10   we list the structural parameters that characterise each empirical law (i.e. effective surface brightness $\mu_e$, effective radius  $r_e$ and $n$-exponent of the Sersic law, and central surface brightness $\mu_0$ and scalelength $r_h$ for the exponential law).}
\end{table*}

\section{Results: the structure of NGC~253}\label{result}

The study of the bar structure is a key issue in addressing the
evolution of self-gravitating disk galaxies. Bars play a major role in
shaping the present properties of disks, since they are responsible
for the re-distribution of the angular momentum and matter within the
disk. The response of a disk to a bar can result in the formation of
pseudo-bulges,   a 3D boxy bar and of substructures such as spiral
  arms and rings \citep[see the review by][]{At12}. In this
framework, we discuss the case of NGC~253 and explore the connections
between the observed substructures in the disk and the orbital
resonances predicted by the disk response to the perturbation by the
bar.

\subsection{The bar structure in NGC~253}\label{bar}
 
The analysis of the light distribution in the shallow J and Ks band images of
NGC~253 (in Sec.~\ref {morph} and \ref {phot}) indicates the presence
of an extended bar   with very bright edges that connect to the
  outer spiral arms (see Fig.~\ref{tileKs}). The isophotal analysis
  (see Fig.~\ref{ellipse}, right panel) suggests that the bar is boxy
  in the inner region and tends to be disky at larger radii, giving a
  typical peanut-shaped end. On average, the bar is redder than the
disk with $1.0 \le J-Ks \le 1.3$~mag   (see Fig.~\ref{profJ_K}).

In this section, we give a more accurate estimate of the length of the
bar and of its strength using the isophote fits, light and color
distributions. Furthermore, we discuss whether the bright features
occurring at the edges of the bar are {\it ansae}, i.e. the typical
symmetric enhancement observed at the end of the stellar bars in many
barred galaxies. 

{\it The estimate of the length of the bar -}   The previous
  estimate of the bar length was about 150~arcsec by
  \citet{Forbes92} from H band images. It was made by measuring the
  bar extension on the image from the center out to the luminous
  edge.  Taking advantage of the very low dust absorption in the Ks
image, we can determine a new and accurate estimate of the deprojected
length of the bar ($l_b$).

According to the method by Gadotti et al. (2007), the observed bar
length ($l_{obs}$) is 1.2 times the radius at which the ellipticity is maximum.  
Given that the ellipticity reaches its maximum at $R=87.5$~arcsec,   $\sim$1.5~kpc, (see
  Fig.~\ref{ellipse}), we estimate $l_{obs} = 104$~arcsec ($\sim$1.7~kpc).  This value is very similar to the preliminary
  estimate of the bar projected length derived by the light profiles,
  given in Sec.~\ref{color} (see also Fig.~\ref{profK}).  The
deprojected bar length $l_b$ is derived by the following relation to
the observed bar length: $l_{obs}=l_b\sqrt{cos^2(\phi) +
  sin^2(\phi)cos^2(i)}$, where $\phi$ is the angle between the bar and
the disk major axis, and {\it i} is the inclination angle. For a 2D
bar, given the angle projected on the sky ($\phi '$), one has
$tan(\phi ') = tan(\phi) cos(i)$. From the isophote fitting, we
estimate $i=74^\circ$ and $\phi ' = 17.2^\circ$, thus giving $\phi =
48.3^\circ$ and $l_b=1.44 \times l_{obs}=151 \pm 12$ arcsec, which is
$\sim$2.5~kpc. The uncertainty on this quantity takes the
errors in the ellipticity and P.A., given by the fit of the isophotes, into account,
which are 0.002 and 4 degrees, respectively.  
%{\bf Because the galaxy is nearly edge-on, the de-projected bar length $l_b$ is larger than the observed length estimated before ($l_{obs} = 104$~arcsec).}

{\it The bar strength -}   Theoretical predictions \citep{At92a,
    At92b} suggest that the degree of dust-lane curvature in barred
  galaxies is inversely proportional to the bar strength, i.e. dust lanes with greater curvature are found in the weaker bars.  This  theoretical expectation was confirmed by observations \citep{Com09,
    Kn02}. For a sample of barred galaxies, with a known value of the
  bar's strength parameter $Q_b$, Knapen et al. (2002) quantified the
  degree of the dust-lane curvature by the ratio $\Delta \alpha $ (in
  unit of degree/kpc), with higher values of $\Delta \alpha $
  corresponding to dust lanes with higher curvature. The diagram of
  $Q_b$ versus $\Delta \alpha $ \citep[see Fig.11 of][]{Kn02} shows a
  clear trend, where higher values of $\Delta \alpha $ correspond to
  lower values of $Q_b$, i.e. weaker bars.  Following the technique
  described by \citet{Kn02}, we estimated the curvature of the dust
  lane in NGC~253. From the deprojected J-Ks color map (shown in 
  Fig.~\ref{colormap}), we measured the change in the angle of the
  tangent to the dust lane. To do this, we have choosen two locations: one
  where the tangent is almost parallel to the bar's major axis and the
  second one where the dust lane curves toward the end of the
  bar. The locations where the dust lanes connect to the nuclear
  regions are excluded since the curvature changes too abruptly.  For
  NGC~253 we derived $\Delta \alpha \sim 25$ degree/kpc. When compared
  with the barred galaxies in the sample studied by Knapen et
  al. (2002), the measured value is in the typical range for weak
  bars, which are characterized by the strength parameter $Q_b \le
  0.2$ \citep[see Fig.~11 of][]{Kn02}.

{\it The edges of the bar -} The J and Ks images of NGC~253 show
round-like and luminous blobs at the ends of the bar (see Fig.~\ref{tileKs}). These are also evident as
high-frequency substructures in the unsharp masked Ks image
(Fig.~\ref{FM}).  When fitting the surface brightness profiles in the
Ks-band along the bar major axis (P.A.=72 degrees), we find that the
peak flux at the edges of the bar is symmetrically located at $R \sim
90$ arcsec ($\sim$1.5~kpc) and is 0.5 mag brighter than the average
bar surface brightness (see Fig.~\ref{2dmod}).  The  morphology and light distribution of these luminous blobs resemble
  those typically observed for the ansae in other galaxies \citep[see Fig.~1 and Fig.~2 in][]{MV07}.

These regions are redder $(J-Ks) \sim 1.2$~mag than the average J-Ks color
along the bar major axis (see Fig.~\ref{profJ_K}): this
observational evidence and the weak bar in NGC~253 are not consistent
with the {\it ansae}, since the latter do not show any color
enhancements and appear mostly in strong bars \citep{MV07}.  An
alternative explanation comes from the $H\alpha$ map of NGC~253
\citep[see Fig.~2 of][]{Hoo96}, which support the identification of
these bright regions with area of star formation. The redder colors
are well explained by the dust absorption.

\subsection{Rings in NGC~253: the role of the orbital resonances}\label{bar_kin}

The high angular resolution and the large field-of-view of the VISTA
images, together with lower dust absorption in the J and Ks-band,
provide robust proof of the existence of two rings within the disk of
NGC~253.   There is a nuclear ring with a radius of about 15 arcsec
  ($\sim$0.2~kpc); it is not uniform, with several bright knots,
  mostly concentrated on the NE side (see Fig.~\ref{zoom}). The
  brightest peak of the whole galaxy light, in the Ks band, is located
  on the SW side of this structure, at $\sim$5.5~arcsec ($\sim0.92$~kpc) far from the
  kinematic center of the galaxy (see Sec.~\ref{morph}). The offset
  between the kinematic and photometric center has also been observed
  in other barred galaxies hosting a nuclear ring \citep{Maz11}, and it varies in the range
  $0.01 - 0.2$~kpc. It is used as a test for the shape of the potential, since the degree
  of such a difference is an indication of non-circular motion
  \citep{Franx94}. Thus, in the case of NGC~253, a
  dynamical model is needed to test if such a difference is due to a
  strong deviation from an axisymmetric potential.

A second ring is located in the main disk enclosing the bar (see
Fig.~\ref{FM}), in the range $158 \le R \le 183$~arcsec ($2.6 \lesssim
R \lesssim 3.1$~kpc).  This component contains active star formation
as suggested by the $H\alpha$ map published by Hoopes et al. (1996),
(see Fig. 2 of their paper).

We now discuss the origin of these two rings in turn.  Rings within
the central kiloparsec are frequently observed in normal disks and
barred galaxies \citep{BC93, BC96, Kn05, Com10} and are most 
likely associated with resonance orbits. In barred galaxies, the bar
plays a crucial role in the redistribution of the angular
momentum. The gas is pushed into orbits near dynamical resonances by
the bar's torque and the star formation is triggered by the high gas
density in these regions \citep[see][for a review]{At12}.  For barred
spiral galaxies with a bar pattern speed $\Omega_b$ and an angular
velocity $\Omega(R)$, there are two basic resonance regions at
different distances from the galaxy center \citep{Lid74}.  These are
the Inner Lindblad Resonance (ILR), where $\Omega (R_{ILR}) = \Omega_b
- K/2$, and K is the epicyclic frequency, and the Outer Lindblad
resonance (OLR) where $\Omega (R_{OLR}) = \Omega_b + K/2$. Rings can be
formed in the vicinity of the ILRs and at the Ultra-Harmonic Resonance
(UHR), where $\Omega (R_{UHR}) = \Omega_b - K/4$ \citep{BC96}. The
radius where the UHR occurs lies between the ILR and the corotation
radius, for which $\Omega (R_{CR}) = \Omega_b$.

\begin{figure}[!ht] 
%\centering
\includegraphics[width=9cm]{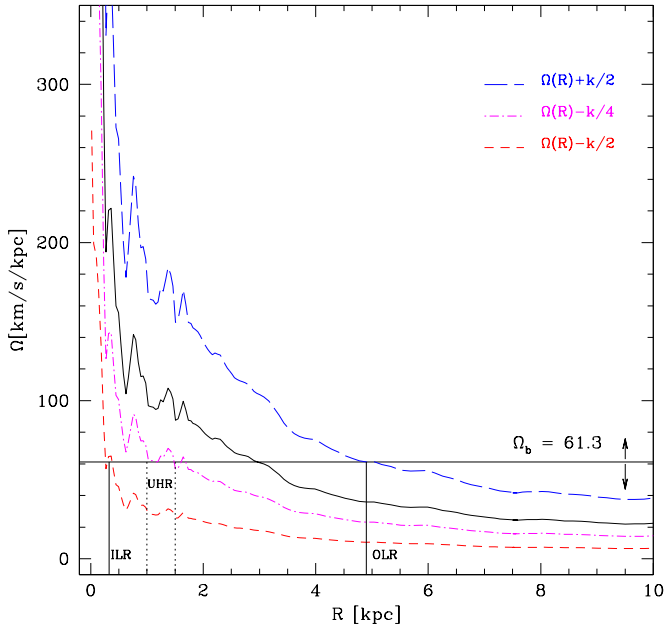}
\caption{Angular velocity curve for the disk of NGC~253   from
    \citet{AM95}: $\Omega(R)$ (solid black curve) and the curves
  relative to $\Omega(R) -K(R)/2$ (short-dashed red line), $\Omega(R)
  + K(R)/2$ (long-dashed blue line) and $\Omega(R) -K(R)/4$
  (long-dashed-point magenta line). The horizontal line corresponds to
  the value of the bar pattern speed $\Omega_b \simeq 61.3$
  km/s/kpc. The two vertical segments correspond to the ILR (at $R
  \sim 0.33$ kpc) and OLR (at $R \sim 4.9$ kpc) radii. The two dotted
  vertical segments correspond to the range of radii where the UHR
  lies at $R \sim 1.3$ kpc. The two vertical arrows indicates the
  lower and upper limit for the $\Omega_b$.}
              \label{omega}
   \end{figure}

In NGC~253 we can estimate the radii at which the resonances occurs by
using the rotation curve along the disk major axis measured by
\citet{AM95}, and verify whether the nuclear and second ring
correspond to any of them.    A first attempt was done by
  \citet{AM95}, based on the previous and very uncertain bar length
  \citep{Forbes92}, that derived a CR radius $R\sim 4.5$~kpc, an
  $\Omega_b = 48$~km/s/kpc, and thus predicted an ILR at about 1~kpc
  and an OLR outside the optical disk.  We accurately measured ($\sim 10\%$) the
intrinsic bar length in Sec.~\ref{bar} to be $l_b=151 \pm 12$ arcsec
$\simeq 2.5$ kpc.  From $l_b$, the radius of corotation $R_{CR}$ is
then $R_{CR}= 1.2\times l_b = 181$ arcsec, $\sim$3~kpc,
\citep{CG89}. Contrary to the measured bar length, this quantity has
much larger uncertainty, $\sim$20$\%$, thus, including also the error
on $l_b$, $R_{CR}$ has a total uncertainty of about 50~arcsec ($\sim0.8$~kpc).  The
measured circular velocity at $R_{CR}$ is $V_{circ} = 183.8$ km/s
\citep{AM95}, hence the bar pattern speed is $\Omega_b =V_{circ}
(R_{CR})/ R_{CR} \simeq 61.3$ km/s/kpc. In Fig.~\ref{omega} we plot
the angular velocity in the disk of NGC~253 $\Omega (R)$ and the
curves relative to $\Omega(R) -K(R)/2 $, $\Omega(R) + K(R)/2$ and
$\Omega(R) -K(R)/4 $.
An estimate of the corresponding radius for ILR and OLR are given by
the loci where $\Omega_b$ insects the $\Omega(R) -K(R)/2 $ and
$\Omega(R) + K(R)/2$ curves, respectively. We find that the ILR falls
at $0.3 \le R \le 0.4 $ kpc and the OLR at $R \sim 4.9$ kpc. The curve
relative to $\Omega(R) -K(R)/4 $ is also added to the plot, and we
estimate the location of the UHR at $1 \le R \le 1.5$ kpc. Taking into
account the error estimate on $R_{CR}$, the $\Omega_b$ could vary
from $\Omega_b^{MIN} =44$ km/s/kpc up to $\Omega_b^{MAX} = 76$
km/s/kpc, as lower and upper limits, respectively. This implies that the
uncertainty on the ILR is about 0.1 kpc. The OLR could fall at
$\sim$4~kpc, for $\Omega_b = \Omega_b^{MAX}$, and at $\sim$7~kpc, for
$\Omega_b = \Omega_b^{MIN}$. Finally, the UHR could be in the range
$0.6 \le R \le 0.9$ kpc, for $\Omega_b = \Omega_b^{MAX}$, and at
$\sim$2.4~kpc, for $\Omega_b = \Omega_b^{MIN}$.  The estimated radius
for the ILR is comparable with that of the nuclear ring, within the
errors.

The OLR is also inside the optical disk, which extends out to 14~kpc
(see Sec. \ref{phot}   and Fig.~\ref{tileKs}).  The presence of the
OLR at $R\sim4.8$~arcmin ($\sim$4.9~kpc) is   very close to the peak of the HI surface density observed in the radial range $2.5 \lesssim R \lesssim 4.5$ arcmin ($\sim 2.5 - 4.5$~kpc)  \citep{P91}. Both the dynamics of
spiral patterns and simulations of bar perturbation predict that gas
piles up at the OLR \citep[see e.g.][]{B96}, and this prediction has
been confirmed by observations of other spiral galaxies
\citep{CA01}. This observational evidence further confirms the
reliability of the OLR estimate from the bar pattern speed $\Omega_b = 61.3$ km/s/kpc.

We now evaluate whether the ring  at $\sim$2.9~kpc,   enclosing the bar, may also originate from orbital resonances. The ring is located very close to where CR lies. However we do not expect a ring
to form at corotation, as stellar orbits do not accumulate at this
radius \citep{Lid74}. We   may also consider this second ring as
originating from the UHR radius, however the latter occurs at a smaller
radius $1 \le R \le 1.5$ kpc, even considering the large uncertainties
on $R_{CR}$ and thus on $\Omega_b$, that gives a maximum value for UHR
at 2.4 kpc.  Thus, bar perturbation theory seems to exclude a
resonance origin for this ring. An alternative origin for this ring may be a merger event or a transient structure formed during
an intermediate stage of bar formation   \citep[see][and reference
    therein]{Com2013}. Simulations by \citet{Atha97} show that the
impact of a small satellite on a barred spiral galaxy can generate a
pseudo-ring that encloses the bar. A merging event has already been
considered as a possible explanation for the extra-planar distribution
of HI gas in NGC~253 \citep{Boo05} and its connection to the $H\alpha$
and X-Ray emission (see Sec.~\ref{intro}).  The resolved stellar
population studies of the deep VISTA data for NGC~253 show a disturbed
disk with extra-planar stars and substructures in the inner halo, also
supporting a possible merger event \citep[][and reference
  therein]{Greggio}.

Alternatively, the bar in NGC~253 could still be forming, and we would
be looking at an intermediate phase in the evolution of the bar. The
N-body simulations that study the role of gas accretion on bar
formation and renewal can also account for the formation of rings
surrounding the bar \citep{BC02}. These are only speculative
suggestions that need to be tested by detailed dynamical models of bar
and disk evolution.

\subsection{The steep outer disk profile in NGC~253}\label{disk}

The most extended component in the shallow images of NGC~253 is the
main disk   that dominates the J and Ks light out to 1034~arcsec
  ($\sim$17~kpc) and 830~arcsec ($\sim$14~kpc) respectively. In the
  deep J-band images, the disk extends further out to 1305~arcsec
  ($\sim$22~kpc), see Fig.~\ref{fit_log}.

The light distribution in the disk is characterized by a down-bending 
Type~II profile, with a break at $R_{br} \simeq 554$~arcsec, i.e. $\sim$9.3~kpc (see Sec.~\ref{ellfit} and the right
panel of Fig.~\ref{fit_log}).  The Type~II down-bending surface
brightness profiles are frequently observed in normal and barred
galaxies, where the radial slope may deviate from a pure exponential
decline beyond 3-5 disk scale lengths (van der Kruit 1979; Pohlen et
al. 2002; Erwin et al. 2008; Pohlen \& Truijllo 2006; Munoz-Mateos et
al. 2013). For NGC~253 the break is observed at about 3 times the
scale length of the inner disk, precisely $R^{Ks}_{br}/r^{in}_h= 3.20
\pm 0.02$ in the Ks band, and $R^{J}_{br}/r^{in}_h= 2.87 \pm 0.03$ in
the J band.

Two possible mechanisms are proposed to explain the observed outer
break in the disks of galaxies: an angular momentum exchange and a
threshold in star formation.  Theoretical studies also showed that
bars can generate such disk breaks \citep{PfF91,Deb06,Foy08}, and in
such cases, the break resides within the OLR. Observations are in
good agreement with the above simulations, showing that the location
of the break radius is consistent with the location of the estimated
OLR \citep{Erw08, MM13} in many disks with down-bending profiles.

\citet{PhT06} investigated the location of the break in a sample of
spiral galaxies, by dividing the Type~II class of profiles into two
subcategories, classical or OLR break (named Type~II-CT and
Type~II-OLR, respectively), according to the physical origin of the
break.  They found that for Type~II-CT  the break radius is, on average, 
at $9.2 \pm 2.4$~kpc, and spans a range from 5.1~kpc to 14.7~kpc. For 
the Type~II-OLR the location of the break spans a wider range,
from 2.4~kpc to 25~kpc, with an average value of the break radius of
$9.5 \pm 6.5$~kpc.  \citet{MM13} recently investigated the connection
between the bars and the location of breaks. In general, most breaks
can be found anywhere in the range of surface brightness $\mu_{br}
\sim 22 - 25$ AB mag arcsec$^{-2}$ or, equivalently, $\Sigma \sim 5
\times 10^7 - 10^8$ $M_{\sun}$ kpc$^{-2}$. They also found that the
range of possible break-to-bar radii $R_{br}/R_{bar}$ is a function of
the total stellar mass, i.e. the most massive disks ($M \ge 10^{10}
M_{\odot}$) have $R_{br}/R_{bar} \sim 2 - 3$, consistent with the
range of OLRs for these galaxies. Galaxies with larger break radii
tend to host weaker bars (i.e. with $\epsilon \le 0.5$). 

By studying the dependence of the star formation rate (SFR) on the
density and dynamics of the interstellar gas in disks, \citet{K89}
proposed a different mechanism for the down bending profile in
exponential disks. The break sets in because of a critical radius
where star formation is inhibited, with the result that a visible
change in slope or truncation in luminosity profile is generated at
this distance from the galaxy center. The critical radius for star
formation can be estimated by the break in the H$\alpha$ surface
brightness profile \citep{MK01} and can be compared with the break
radius $R_{br}$ estimated from the light profiles of the underlying
stellar population.

What is a realistic explanation for the observed break in the disk of
NGC~253? On the basis of the joint analysis of the new NIR VISTA and
previous kinematical data for the NGC~253 disk, we place the OLR of
the bar at $R \sim 4.9$~arcmin ($\sim$4.9~kpc). Taking into account
the large uncertainties on $R_{CR}$ and thus on $\Omega_b$, the
maximum value for OLR radius is $\sim$7~kpc.    The break radius in
  the light profiles of NGC~253 is at $\sim$9.3~kpc. Even if this
  value is comparable with both the estimates derived by \citet{PhT06}
  for Type~II-CT and Type~II-OLR profiles, it is much further out than
  the location of the OLR.   This suggests that the angular momentum
exchange, driven by the bar, may not be the origin of the down-bending
profile for the disk of NGC~253. This is further confirmed by the
ratio $R_{br}/R_{bar} \sim 6$ for NGC~253, which is more than a factor
of two greater than the range of values estimated by Munoz-Mateos et
al. (2013) for galaxies with breaks in their light profiles consistent
with the OLRs.
On the other hand, considering the average deprojected ellipticity of
the bar in NGC~253   ($\sim$0.55, see Fig.~\ref{ellipse}), the
estimated break radius turns out to be consistent with the location of
the ones measured in low-mass disk galaxies, with large break radii
and hosting weak bars. This is consistent with previous observational
studies indicating that ``classical breaks'' at larger radii are more
common in late-type disk galaxies, while the ''OLR breaks" are more
frequent in early-type disks \citep{PhT06, Erw08, MM13}.

On the basis of the above discussion, it is unlikely that the orbit
resonances are responsible for the observed break in the surface
brightness profile of the outer disk in NGC~253. Hence we consider
the alternative origin due to a threshold in the star formation. In
NGC~253, the H$\alpha$ surface brightness profile is published by
Hoopes et al. (1996; see Fig. 6 of that paper). For $9 \le R \le
10$~arcmin ($9.1 \le R \le 10.1$~kpc) the slope changes and shows a
sharper decline with respect to smaller radii. The observed break in
the H$\alpha$ profile falls in the same range where the break in the
NIR light profiles are observed. This suggests that some mechanism has
inhibited the star formation in the disk of NGC~253 from this radius
outward. Both the H$\alpha$ and the HI distribution are less extended
than the stellar counterpart, reaching a distance from the center of
galaxy of $\sim$11~kpc \citep{Boo05}. In particular, the deep data in
the J band show that the "outer disk" is   a factor of two more
extended than the H$\alpha$ and HI, reaching a distance of
$\sim$22~kpc from the center (see Fig.~\ref{fit_log}). This is also
confirmed in deep optical band observations by \citet{MH97}. As
  shown by \citet{Greggio}, the halo of the galaxy starts from about
  20~arcmin ($\sim 20.2$~kpc).
The HI distribution \citep{P91} exhibits a peak in the range $2.5
\lesssim R \lesssim 4.5$~arcmin ($2.53 \le R \le 4.54$~kpc). At this
HI peak, the neutral gas surface density also has a peak of $\Sigma
\sim 7$ M$_\odot pc^{-2}$, from which it then declines to a value of
$\Sigma \sim 2.5$ M$_\odot pc^{-2}$ at the break radius $R_{br} = 9.3$
kpc. This value is significantly lower than the critical surface
density at this radius, which is $\Sigma \sim 9.7$ M$_\odot pc^{-2}$
according to the Kennicutt law \citep{K89, MK01}.  This clearly
suggests that some past mechanism has   redistributed the gaseous
disk and subsequently inhibited the star formation.  As already
suggested by Boomsma et al. (2005), there are several mechanisms that
may be responsible for the truncation of the gaseous disk in NGC~253
and the consequent inhibition of the star formation: the gas in the
outer layers may have been ionized by the hot stars and starburst in
the disk of NGC~253; the gas may have been removed by the ram pressure
stripping of other galaxies in the Sculptor group or, finally, a
merger event may have led to significant disturbances in the disk and halo.

\section{Summary }\label{concl}

We have performed a detailed photometric analysis   of the barred spiral
  galaxy NGC~253 from the deep and shallow data in the J band and
  shallow data in the Ks band, taken during the Science Verification
run of the new VISTA telescope on Paranal.  The disk,   which
  extends out to 22~kpc in the deep J band image, hosts three
prominent features (see Figure~\ref{tileKs}): {\it i)} the
bright and almost round bulge with a diameter of about 1 kpc; {\it
  ii)} the bar,   up to 1.7~kpc, with a typical peanut shape
ending in very bright edges; and {\it iii)} the spiral arms which
dominate in the regions of the disk. In addition to the average
light distribution of these main components, the following two
substructures are observed (see Fig.~\ref{FM} and Fig.~\ref{zoom}): a nuclear ring with a diameter of  $\sim$0.5~kpc; and a second ring   enclosing to the bar with an average radius of $\sim$2.9~kpc located in the disk. All the above components are already visible in the J-band image (Fig.~\ref{tileJ}), but are more clearly identified in the Ks
image (see Fig.~\ref{tileKs}). In Table~\ref{tab_n253} we list the
main parameters for the structural components in the disk of NGC~253
(radial extension, ellipticity, P.A., and average colors), based on our photometric analysis.

\begin{table}
\caption{Photometric parameters, derived by the analysis in the Ks
  band, which characterize the main components and substructures in
  NGC~253.}\label{tab_n253}
  \centering
\begin{tabular}{l c c c c}     % 6 columns 
\hline\hline     
Main components in NGC~253 \\                    
\hline                  
  & $length$ & $\epsilon$ & P.A. & J-Ks\\
  & kpc & & degree & mag\\
\hline                  
bulge & 0.6 & 0.4 & 58 & 1 - 2\\
bar & 1.7 & 0.4 - 0.7 & 72 & 1 -1.3 \\
inner disk & 2.9 & 0.7 - 0.8 & 52 & 0.8 - 1 \\
outer disk & 1.5 & 0.8 & 52 & 0.3 - 0.8 \\
\hline\hline                  
substructures in NGC~253 \\                  
\hline                  
nuclear ring & 0.2 & 0.6 & 52 & 1.5 - 1.8\\
ring & 2.9 & 0.8 & 52 & 1.2\\
\hline                  
\end{tabular}
\tablefoot{  The first column lists the different
    structures in NGC~253. In the Col.~2, are the observed length, for
    the disk numbers refers to the scale length. In Col. 3 and 4 are
    the ellipticity and P.A., respectively, derived by the fit of the
    isophotes. In Col.~5 are the minimum and maximum values of the
    J-Ks color.}
\end{table}

On the basis of the results from the quantitative photometry carried
out in this study, we have described the structure of the nucleus, bar and
disk, the connection between the observed substructures in the disk
and the Lindblad resonances predicted by the bar/disk kinematics
(Sec.~\ref{result}). The main results of this analysis are:

\begin{itemize}

\item   from the degree of the curvature of the dust-lane in the J-Ks
  color map, we obtain an indication for a weak bar in NGC~253 (see
  Sec.~\ref{bar}). Since the bright knots at the end of the bar are
  redder than the inner regions, and also given the late-type
  morphology and strength of the bar, these bright regions are not the
  ansae typically observed in other barred galaxies, but rather they
  are regions of local star formation (see Sec.~\ref{bar});

\item from the measurement of the bar's   deprojected length on the
  new Ks image, we derive a new value for the corotation radius (CR)
  in NGC~253, located at $R_{CR} \sim 3$~kpc. We
  then estimate the bar pattern speed, $\Omega = 61.3$ km/s/kpc, and
  the corresponding radii for LRs.  We find that the ILR is at $0.3
  \le R \le 0.4$~kpc, the OLR is at $R \sim 4.9$~kpc, and the UHR is in the
  range $ 1 \le R \le 1.5$~kpc;

\item   the nuclear ring observed in NGC~253 is located at the
  ILR. Its morphology and radius is similar to those found for other
  nuclear rings with a resonant origin observed in several other
  barred galaxies \citep{BC93, Com10, Maz11}; 

\item the presence of the OLR at $R \sim 4.9$~kpc is
  consistent with the peak of the HI surface density observed at
  similar radii. We cannot associate the ring at 2.9 kpc with UHR,
  which is expected at smaller radii.  The ring may, in fact, be the
  result of a minor merger event or, alternatively, a transient
  structure formed during an intermediate stage of bar formation;

\item   the disk of NGC~253 has a down-bending profile with a break
  at $R \sim$9.3~kpc, which corresponds to about 3 times the
  scale length of the inner disk. We conclude that such break may most
  likely arise from a threshold in star formation. The exact mechanism
  which is responsible for the truncation of the gaseous disk in
  NGC~253, and the consequent inhibition of the star formation, is
  still to be identified. Three possible mechanisms are presented:
  ionization of the gas in the outer layers by the hot stars and
  starburst in NGC~253, ram pressure stripping of the gas by the other
  galaxies in the Sculptor group or by a merger event.

\end{itemize}

  A merger event has been invoked several times already and by independent authors to explain the current morphology of NGC~253 \citep{Boo05}. 
This may in fact explain the presence of an extended asymmetrical stellar halo plus a Southern spur, in the deep optical images and recently confirmed by the new deep VISTA data \citep{Greggio}, the HI off-plane plume, elongated perpendicular to the disk major axis, the truncation of the gaseous disk and the consequent inhibition of the star formation and the presence of the ring at the barÕs edges, that is clearly detected in the new VISTA shallow images.

To conclude, the new VISTA imaging data presented in this paper provide a detailed and more complete view of NGC~253, and illustrate the amazing capability of the VISTA telescope for studies that require high angular resolution   on a large field of view.

%---------------------------------------------------------

\begin{acknowledgements}
This work is based on observations taken at the ESO La Silla Paranal
Observatory within the VISTA Science Verification Program
ID~60.A-9285(A).  We are very grateful to the referee, Michael Pohlen,
for his comments and suggestions that improved this work.  We thank
Jorge Melnick for initiating the VISTA Science Verification and its
organization, Thomas Szeifert and Monika Petr-Gotzens for the
assistance and help during the observing run, and Jim Lewis, Simon
Hodgkin and Eduardo Gonzalez-Solares from CASU for their expert
contribution to the VISTA data processing.  E.I wish to thank ESO for
the financial support and hospitality given during her several visits
in 2011 and 2012 to work on the SV data. The authors wish to thank
M. Capaccioli, E.M. Corsini, T. de Zeeuw, E. Emsellem, K.C. Freeman
and O. Gerhard for useful comments and discussions.

\end{acknowledgements}

%\begin{thebibliography}{}
\bibliographystyle{aa}
\bibliography{NGC253.bib}

\end{document}